\documentclass{aa}

\usepackage{graphicx}
\usepackage{txfonts}
\usepackage{color}
\usepackage{longtable}
\usepackage{hyperref}

\newcommand{\tablenotea}[1]{\parbox{18.4cm}{\indent \footnotesize{#1}}}

\begin{document}

\title{The mutual influence of disequilibrium composition and temperature in exoplanet atmospheres}

\titlerunning{The mutual influence of disequilibrium composition and temperature in exoplanet atmospheres}
\authorrunning{Ag\'undez}

\author{Marcelino~Ag\'undez}

\institute{
Instituto de F\'isica Fundamental, CSIC, Calle Serrano 123, E-28006 Madrid, Spain\\ \email{marcelino.agundez@csic.es} 
}

\date{Received; accepted}
 
 \abstract
{Astronomical observations have provided an extensive body of evidence for the existence of disequilibrium chemistry in many exoplanet atmospheres, and this departure from a chemical equilibrium composition may have an impact on the temperature of the atmosphere itself. We have developed a 1D atmosphere model that solves in a self-consistent manner the evolution of temperature and disequilibrium chemistry in the vertical direction. The temperature is solved in radiative-convective equilibrium and the disequilibrium composition is computed including thermochemical kinetics, photochemistry, and vertical mixing. Thermochemical kinetics is based on a reaction network built from scratch that includes 164 gaseous species composed of H, C, N, O, S, Si, P, Ti, He, and Ar, connected by 2352 forward reactions. To investigate the mutual influence between disequilibrium chemistry and temperature in exoplanet atmospheres, we have applied our model to the well-known gas giant exoplanets WASP-33b, HD\,209458b, HD\,189733b, GJ\,436b, and GJ\,1214b, which cover different degrees of insolation and metallicity, and to secondary atmospheres that exoplanets characterized in the future may plausibly have. We find that for irradiated gas giants with solar or supersolar metallicity, the corrections to the temperature due to disequilibrium chemistry are relatively small, on the order of 100 K at most, in agreement with previous studies. Although the atmospheric composition of some of these planets deviates significantly from chemical equilibrium, the impact on the temperature is moderate because the abundances of the main atmospheric species that provide opacity, such as H$_2$O, CO$_2$, CO, and/or CH$_4$, are not seriously modified by disequilibrium chemistry. An impact on the temperature greater than 100 K appears in hot Jupiters due to TiO, which is predicted to be seriously depleted by UV photons in the upper layers. However, the extent of this depletion, and thus of its impact on the temperature, is uncertain due to the lack of knowledge about TiO photodestruction. In secondary atmospheres, the impact of disequilibrium chemistry on the temperature depends on the composition. In atmospheres dominated by H$_2$O and/or CO$_2$ the temperature is not affected to an important extent. However, reducing atmospheres dominated by CH$_4$ and oxidizing atmospheres dominated by O$_2$ see their temperature being seriously affected due to the important processing of the atmospheric composition induced by disequilibrium chemistry.}


\maketitle

\section{Introduction}

A variety of processes, such as photochemistry, thermochemical reactions, and mixing, can drive the chemical composition of a planet atmosphere out of chemical equilibrium. A disequilibrium composition can have an impact on the emission and transmission spectrum of the planet, and this can be probed with telescopes such as the James Webb Space Telescope (JWST), Spitzer, the Hubble Space Telescope, or Ariel \citep{Blumenthal2018,Venot2018a,Baxter2021,Roudier2021,Kawashima2021}. Nice recent examples of the imprint of disequilibrium chemistry on the spectra of exoplanets are provided by the detection of photochemically produced SO$_2$ in the atmosphere of the hot Jupiters WASP-39b and WASP-107b \citep{Alderson2023,Rustamkulov2023,Tsai2023,Dyrek2023}, the lack of CH$_4$ on the nightside of WASP-43b \citep{Bell2024}, or the low abundance of CH$_4$ in WASP-107b \citep{Sing2024,Welbanks2024}.

A disequilibrium composition can also have an impact on the temperature of the atmosphere itself, which is to a large extent controlled by the abundances of the main atmospheric constituents that provide opacity at infrared, visible, and ultraviolet (UV) wavelengths. This phenomenon is well known in Earth's atmosphere, where the formation of ozone due to photochemistry originates a temperature inversion at 6-20 km of altitude. Moreover, disequilibrium composition and temperature can be mutually influenced because if the thermal atmospheric structure is modified as a result of disequilibrium chemistry, the composition can in turn be altered in response to the change induced in the temperature. This feedback between disequilibrium composition and temperature was addressed for the warm Neptune GJ\,436b through iterative runs of a 1D radiative-convective code and a 1D photochemical kinetics code \citep{Agundez2014a}, where it was found that corrections to the temperature due to disequilibrium were moderate ($<$\,100 K). A similar approach in which a 1D radiative-convective code is coupled to a nonequilibrium chemical kinetics code was adopted by \cite{Tremblin2015} to model brown dwarf atmospheres and by \cite{Drummond2016} to model the atmospheres of the hot Jupiters HD\,209458b and HD\,189733b. These latter authors found that temperature differences due to the inclusion of nonequilibrium chemistry are restricted to less than 100 K and depend on the adopted eddy diffusion coefficient. More recently, \cite{Mukherjee2023,Mukherjee2024,Mukherjee2025a} explored the effect of nonequilibrium chemistry on the atmospheric temperature of gas giants and found that corrections to the temperature, which can be up to $\sim$\,100 K, become more important as the metallicity and the internal temperature increase. A different view of the temperature-composition feedback was brought by \cite{Nicholls2023}, who investigated the impact of stellar flares on the atmospheric temperature and composition, and by \cite{Kecskemethy2024}, who found that the evolution in the internal heat of the planet affects the composition in H$_2$-dominated atmospheres.

Here we investigate the mutual influence of disequilibrium composition and temperature in the dayside of exoplanet atmospheres of different degrees of insolation and elemental composition. To that purpose we have developed a 1D self-consistent code in which the evolution of the vertical distribution of temperature (driven by radiation and convection) and disequilibrium composition (driven by thermochemical kinetics, photochemistry, and vertical mixing) is solved as a function of time. We describe the numerical code in Sect.\,\ref{sec:code}, introduce the exoplanet atmospheres investigated in Sect.\,\ref{sec:models}, present the results of the models in Sect.\,\ref{sec:results}, discuss the implications in Sect.\,\ref{sec:discussion}, and summarize the main conclusions in Sect.\,\ref{sec:conclusions}.

\section{Numerical code} \label{sec:code}

We developed a 1D numerical code written in Fortran aiming to describe in a self-consistent way the nonequilibrium chemical composition and temperature along a vertical column of planetary atmosphere. The code, which is publicly available (see Sect.\ref{sec:data}), solves the vertical distribution of temperature and composition as a function of time, starting from some initial state. The initial chemical composition can be either specified as an input, in which case the initial temperature is computed for that particular composition, or calculated under chemical equilibrium given a set of elemental abundances, in which case temperature and chemical equilibrium composition are solved iteratively until the temperature varies by less than 1 K between successive iterations. Once the initial state is established, the code solves the nonequilibrium chemical composition along the vertical column of atmosphere as a function of time, and the temperature is evaluated at different intermediate times according to the true actual composition. The first intermediate time at which temperature is evaluated is 10$^{-10}$ s, and then intermediate times increase linearly on a logarithmic scale, with a time step of a factor of 10$^1$ until the integration time reaches 8.64\,$\times$\,10$^4$ s (one day) and a factor of 10$^{0.2}$ later on to more closely follow the feedback between composition and temperature. This time step scheme is a good compromise between accuracy and computation time. Reducing the time step leads to temperature changes that are usually smaller than 1 K. The integration with time continues until a steady state is reached, which we consider to happen when the time is greater than one day and the relative abundance variation\footnote{The relative abundance variation at an intermediate time, $t_k$, with respect to the previous intermediate time, $t_{k-1}$, was calculated as the maximum value of the quantity $|f_i^j(t_k) - f_i^j(t_{k-1})|/f_i^j(t_{k-1})$, where $f_i^j$ is the mixing ratio of species $i$ in layer $j$ and the indexes $i$ and $j$ run for all mixing ratios, $f_i^j$, with non-negligible values, above 10$^{-20}$.} reaches a value below 10$^{-2}$ (or below 10$^{-1}$ during three consecutive intermediate times). This convergence criterion is similar to that adopted in other chemical kinetics codes used to model planetary atmospheres, such as VULCAN \citep{Tsai2017} or photochem \citep{Wogan2024,Mukherjee2025a}, where the maximum relative abundance variation adopted is in the range (1-5)\,$\times$\,10$^{-2}$. A steady state is typically reached after (1-100)\,$\times$\,10$^7$ s in hot atmospheres ($>$\,1000 K), while in cooler atmospheres it may take somewhat longer, up to 10$^{11}$ s.

The code is organized into several modules that are called every time they are needed. The main module is the one in charge of solving the nonequilibrium chemical composition as a function of time. There is a module that computes the chemical equilibrium composition in case this is required for the initial state, while a radiative-convective module calculates the vertical distribution of the temperature for a given chemical composition at the initial time and at every intermediate time.

\subsection{Nonequilibrium chemistry module}

The core of this module is the one used in \cite{Agundez2014b} to model the evolution in the chemical composition of a vertical column of atmosphere, except that here we stay 1D and not pseudo-2D. That is, the physical conditions of the vertical atmosphere column remain static, unlike in the pseudo-2D approach in which they vary with time following the rotation around the planet equator. The evolution of the vertical distribution of a species, $i$, is described by the coupled continuity-transport equation
\begin{equation}
\frac{\partial f_i}{\partial t} = \frac{P_i}{n} - f_i L_i - \frac{1}{n r^2} \frac{\partial (r^2 \phi_i)}{\partial r}. \label{eq:continuity}
\end{equation}
In the above equation, $f_i$ is the mixing ratio of species $i$, $t$ is the time, $n$ is the total number of particles of any species per unit volume, $r$ is the radial distance to the center of the planet, $P_i$ and $L_i$ are the rates of production and loss, respectively, of species $i$, and $\phi_i$ is the vertical transport flux of particles of species $i$ (positive upward and negative downward). The first two terms on the right side of Eq.~(\ref{eq:continuity}) account for the formation and destruction of species $i$ by chemical and photochemical processes, while the third term describes vertical transport in a spherical atmosphere.

In the practice, the atmosphere is divided into a finite number of layers (typically 70) and the continuous variables in Eq.~(\ref{eq:continuity}) are discretized as a function of altitude. After the discretization, Eq.~(\ref{eq:continuity}) reads
\begin{equation}
\frac{\partial f_i^j}{\partial t} = \frac{P_i^j}{n^j} - f_i^j L_i^j - \frac{\big(r^{j-1/2}\big)^2 \phi_i^{j-1/2} - \big(r^{j+1/2}\big)^2 \phi_i^{j+1/2}}{n^j \big(r^j\big)^2 \big(z^{j-1/2} - z^{j+1/2}\big)}, \label{eq:continuity-discrete}
\end{equation}
where $z$ is the altitude in the atmosphere with respect to a reference level that corresponds to the radius of the planet (arbitrarily set to 1 bar) and the superscript $j$ refers to the $j^{\rm th}$ layer, while $j+1/2$ and $j-1/2$ refer to its lower and upper boundaries, respectively, so that layers are ordered from top to bottom. The nonlinear system of first-order ordinary differential equations given by Eq.\,(\ref{eq:continuity-discrete}) is integrated as a function of time using the backward differentiation formula implicit method for stiff problems implemented in the Fortran solver DLSODES, included in the ODEPACK package\footnote{\tiny \url{https://computing.llnl.gov/projects/odepack}} \citep{Hindmarsh1983,Radhakrishnan1993}.

Equation~(\ref{eq:continuity}) contains the three main processes that shape the distribution of species in a vertical column of atmosphere, which are thermochemical kinetics, photochemistry, and vertical mixing. Hereafter we describe how the code deals with each of them.

\subsubsection{Thermochemical kinetics}

The contribution of thermochemical kinetics to the production and loss terms in Eq.\,(\ref{eq:continuity-discrete}) is taken over by gas-phase chemical reactions. We use the term “thermochemical kinetics” instead of simply “chemical kinetics” because for each reaction included in the reaction network with a given rate coefficient (which we name “forward reaction”) we also include the reverse reaction with the rate coefficient calculated via detailed balance using the thermochemical properties of the species involved. We have built from scratch a reaction network involving the elements H, C, N, O, S, Si, P, Ti, He, and Ar, which contains 164 species connected by 2352 (forward) reactions. The species considered are given in Table\,\ref{tab:species}. All them are gaseous neutral species, and thus the network is suitable for modeling the neutral gaseous layers of planetary atmospheres, but not the uppermost ionosphere or the appearance of aerosols. This is clearly a limitation of the model because clouds (formed by the condensation of refractory elements from the gas phase) and hazes (produced by photochemistry) are likely to occur in some exoplanet atmospheres \citep{Pont2013,Parmentier2013,Helling2023,2024KIE} and they would probably have an effect on the atmospheric temperature. Moreover, we include Si and Ti but species containing other metals, such as Fe, Mg, Al, Na, and K, are likely to be abundant enough in hot atmospheres that they can impact the temperature. A detailed network of reactions involving these metal-bearing species is needed, however, to properly describe their abundances. Currently, the reaction network includes 2067 bimolecular reactions, 281 pressure-dependent reactions, and two dummy reactions involving the chemically inert atoms He and Ar. The rate coefficients have been taken from the chemical kinetics literature, either from experimental studies, theoretical ones, or from compilations. We have made extensive use of the NIST Chemical Kinetics Database\footnote{\tiny \url{https://kinetics.nist.gov/}} during the search for documented reaction rate coefficients. In addition, the network has been cross-checked against published evaluations, compilations, and networks developed for Earth atmosphere chemistry \citep{1997DEM,1997ATK,2004ATK,2006ATK,2020BUR}, combustion chemistry \citep{1984HAN,1986TSA,1987TSA,1988TSA,1990TSAa,1991TSAa,1991TSAb,1992TSAa,1992BAU,1994BAU,GRI-Mech,2000DEA,2001KONa,2005BAU,2013ZHO,2017SON}, exoplanet atmospheres chemistry \citep{2002MOS,2012VEN,2016MOS,2016ZAH,Rimmer2016,2021HOB,2021TSA,2022JAZ,2024KIE}, and interstellar and circumstellar chemistry \citep{2004SMI,2006AGU,2017VID,2017LOI,2024MIL}. The list of reactions and associated rate coefficients, together with their references, are given in Table\,\ref{tab:reactions}.

In the case of pressure-dependent reactions in which there is information on both the three-body direction and the thermal dissociation one, we usually adopted as forward reaction the three-body process and calculated the rate coefficient of the reverse thermal dissociation via a detailed balance using the thermochemical properties of the species involved. Since thermal decomposition reactions are usually measured at high temperatures, the extrapolation to room temperature may result in a rate coefficient for the reverse three-body reaction (calculated via detailed balance) that could be wrong by several orders of magnitude. As an example, if we use the $k_0$ expression measured in the temperature range 4060-6060 K for the reaction CN + M $\rightarrow$ C + N + M, 4.15\,$\times$\,10$^{-10}$ $\exp (-71000/T)$ cm$^3$ s$^{-1}$ \citep{1989Mozzhukhin}, we would obtain a rate coefficient for the reverse reaction that would be many orders of magnitude above the value of 9.4\,$\times$\,10$^{-33}$ cm$^6$ s$^{-1}$ that is documented at 298 K in the NIST database \citep{1974KLE}.

The rate coefficient of bimolecular reactions was parameterized as a function of temperature through the usual modified Arrhenius expression,
\begin{equation}
k = \alpha \, \bigg( \frac{T}{300} \bigg)^\beta \exp{(-\gamma/T)}, \label{eq:k_modified-Arrhenius}
\end{equation}
where $T$ is the temperature in degrees Kelvin. In the case of pressure-dependent reactions, the rate coefficient is given by
\begin{equation}
k = \frac{k_0 {\rm [M]} k_{\infty}}{k_0 {\rm [M]} + k_{\infty}} \, F = k_{\infty} \bigg( \frac{P_r}{1 + P_r} \bigg) \, F, \label{eq:k_p-dependent}
\end{equation}
where $k_0$ and $k_\infty$ are the low- and high-pressure limit rate coefficients, respectively, and $P_r$ is the reduced pressure, defined as $P_r = k_0 {\rm [M]} / k_\infty$. The term [M] is the sum of the volume densities of the different species, $i$, corrected by their corresponding collision efficiencies; that is,
\begin{equation}
{\rm [M]} = \sum_{i=1}^{N_S} f_{coll,i} \, n_i,
\end{equation}
where the sum extends to the total number of species, $N_S$. If the bath gas associated with $k_0$ is known, we adopted collision efficiencies of 1.00 for N$_2$, 0.70 for Ar and He, 1.50 for CO, 2.00 for H$_2$, CO$_2$, and CH$_4$, 6.00 for H$_2$O, and 1.00 for any other species, based on the examination of a dozen of reactions and on calculated values in the GRI-Mech database\footnote{\tiny \url{http://combustion.berkeley.edu/gri-mech/}} \citep{GRI-Mech}. If no information was available on the bath gas, we adopted collision efficiencies of 1.00 for all species for simplicity. In the case of the reaction between CH$_3$ and OH, a modified version of Eq.\,(\ref{eq:k_p-dependent}) based on \cite{2007JASa} was used to take into account the special dependence with pressure. It is often the case that information is only available for $k_0$ or $k_\infty$, but not for both. In those cases, the missing rate coefficient can be crudely estimated using
\begin{equation}
\frac{k_0}{k_\infty} = 10^{-20.8 + 1.6 N_{\rm C}} {\rm cm^3} \Bigg( \frac{T}{300} \Bigg)^{-2}, \label{eq:k0_kinf}
\end{equation}
where $N_{\rm C}$ is the number of carbon atoms involved in the reaction. Eq.\,(\ref{eq:k0_kinf}) was inspired by \cite{2012VUI} and was empirically obtained after examination of more than 80 reactions with $N_{\rm C}$\,=\,0-6 for which both $k_0$ and $k_\infty$ were known.

The parameter $F$ in Eq.\,(\ref{eq:k_p-dependent}), which accounts for the behavior of $k$ in the fall-off region where ${\rm [M]} \sim k_\infty / k_0$, was evaluated using the Troe formalism \citep{Gilbert1983}
\begin{equation}
\log F = \frac{\log F_c}{1 + \Big( \frac{\log P_r + c}{N - d [\log P_r + c]}\Big)^2},
\end{equation}
where $c = -0.4 - 0.67 \log F_c$, $N = 0.75 -1.27 \log F_c$, $d = 0.14$, and $F_c$ is the broadening factor. We used a simple linear dependence of $F_c$ with temperature, $F_c = A + B \times T$. \cite{1992TSAa} argued that the error introduced by this simple dependence with respect to more complex expressions is no more than a factor of 1.20. To avoid unreasonable values of $F_c$, we did not extrapolate outside the temperature range of validity of the fitting parameters A and B. If the parameter $F_c$ is not known, one could adopt the Lindemann-Hinshelwood formalism, in which $F=1$, although we adopted $F_c=0.5$ independent of temperature, which is the average value found after an examination of more than 50 reactions.

The parameters $\alpha$, $\beta$, and $\gamma$ in Eq.\,(\ref{eq:k_modified-Arrhenius}), which describe the dependence on the temperature of the rate coefficient, $k$, of bimolecular reactions, and of $k_0$ and $k_\infty$ for pressure-dependent reactions, are strictly valid only over a given temperature range, although in the practice the expressions are extrapolated in temperature as needed. We examined the behavior of the rate coefficients of all reactions over the temperature range 100-4000 K and found that for some reactions the extrapolation in temperature results in abnormally high values. To prevent this unchemical behavior, we considered upper limits to reaction rate coefficients. For bimolecular reactions (with units of cubic centimeter per second, this also includes $k_0$ for thermal dissociations and $k_\infty$ for three-body reactions) we set $k \le 2 \times10^{-9}$ cm$^3$ s$^{-1}$, which is at the high edge of collision rate coefficients for neutral-neutral reactions. In the case of termolecular reactions, we set $\log k \le -29.5 + 2.15 N_{\rm C}$, where $k$ has units of centimeter to the sixth power per second and $N_{\rm C}$ is the number of C atoms involved in the reaction. This latter expression was again inspired by \cite{2012VUI} and was derived empirically after examination of more than 150 three-body associations. We divided the reactions into three categories, A, B, and C, according to the uncertainty we assign for the rate coefficient. For those reactions studied experimentally in which the dependence with temperature was known or at least understood, we assigned an uncertainty lower than a factor of two (type A), while an uncertainty between a factor of two and a factor of ten (type B) was usually assigned to reactions that had only been studied theoretically, and an error higher than a factor of ten (type C) was assigned to reactions for which the rate coefficient had been estimated or guessed based on the behavior of similar reactions or on chemical intuition.

\subsubsection{Photochemistry}

Photodissociation processes, triggered by the penetration of stellar UV photons into the planet atmosphere, also contribute to the production and loss terms in Eq.\,(\ref{eq:continuity-discrete}). Photoionization processes are not included because we are interested in the neutral layers of exoplanet atmospheres and only neutral species are considered. Ions are relevant in the uppermost atmospheric layers \citep{Lavvas2014,Rimmer2016,Bourgalais2020}. The photodissociation rate of a given species in a certain atmospheric layer depends on the wavelength-dependent UV flux in that layer and the relevant wavelength-dependent cross section of that species. The UV flux is calculated by solving the radiative transfer of stellar UV photons in the vertical direction. The stellar UV spectra adopted for the exoplanet atmospheres modeled here are discussed in Sect.\,\ref{sec:models}. The path length in each layer was computed considering spherical geometry and the zenith angle of incidence of stellar photons measured from the local vertical. Absorption and Rayleigh scattering by molecules and atoms contribute to the attenuation of stellar UV photons. Rayleigh scattering was treated using a simplified two-ray iterative algorithm \citep{Isaksen1977} that allows for a fast computation of the UV radiative transfer. The Rayleigh scattering cross sections were calculated for the most abundant species from their polarizability \citep{Tarafdar1969}. We ensured that all species with mixing ratios above 10$^{-4}$ contribute to Rayleigh scattering.

Photoabsorption and photodissociation cross sections were compiled from original sources in the literature or from databases (references are given in Table\,\ref{tab:species}). In the case of atoms, photoabsorption cross sections were mainly taken from the NORAD-Atomic-Data database\footnote{\tiny \url{https://norad.astronomy.osu.edu/}} \citep{Nahar2020,Nahar2024}, while for molecules we made extensive use of the MPI-Mainz UV/VIS Spectral Atlas of Gaseous Molecules of Atmospheric Interest\footnote{\tiny \url{https://www.uv-vis-spectral-atlas-mainz.org/uvvis/}} \citep{Keller-Rudek2013}, which contains experimental photoabsorption cross sections mainly for stable molecules, the Leiden database of photodissociation and photoionization of astrophysically relevant molecules\footnote{\tiny \url{https://home.strw.leidenuniv.nl/~ewine/photo/}} \citep{vanDishoeck2006,Heays2017,Hrodmarsson2023}, which contains experimental and theoretical photoabsorption and photodissociation cross sections for stable molecules and radicals, and the Photo Rate Coefficient Database\footnote{\tiny \url{https://phidrates.space.swri.edu/}} \citep{Huebner1992}, which contains photoabsorption and photodissociation cross section together with information on branching ratios for the different photodissociation channels for molecules of interest in Solar System atmospheres. A discussion on the cross section data for some of the species included can be found in \cite{Agundez2018}. Experimental studies usually provide information on the photoabsorption cross section, without specific knowledge on whether the absorption of the UV photon leads to dissociation, ionization, or fluorescence. In such cases, we assumed that absorption of UV photons with wavelengths above the photoionization threshold and below the photodissociation threshold only leads to dissociation, with no contribution of fluorescence. This is probably a good approximation for wavelengths below 150 nm, but not for longer wavelengths. Most experimental studies provide the photoabsorption cross section at room temperature, although it is known that at higher temperatures the cross section can increase significantly \citep{Schulz2002,Venot2013,Venot2018b,Grosch2015,Matsugi2016,Pattillo2018}. We do not include temperature-dependent UV cross sections, mainly because of the scarcity of available data. However, the impact of such a temperature dependence on the atmospheric composition, which has been studied to some extent for CO$_2$ \citep{Venot2013,Venot2018b}, is an interesting niche to explore \citep{Fortney2019,Chubb2024}. To ensure that all molecules are destroyed by stellar UV radiation in the uppermost atmospheric layers, we considered that all molecules included have some photodissociation process. In the case of those molecules for which there are no cross section data, we assumed a guess value of 1 Mb\footnote{1 Mb is equal to 10$^{-18}$ cm$^{-2}$.} from the photoionization threshold to 250 nm, or in the range of 100-250 nm if the ionization threshold was not known (see Table\,\ref{tab:species}). As in the case of reaction rate coefficients, we divided the UV cross section data into three categories, A, B, and C, according to their uncertainty. We assigned an uncertainty lower than a factor of two (type A) when experimental data was available over a sizable wavelength range, between a factor of two and factor of ten (type B) when there were only theoretical data or experimental data over a limited wavelength range, and higher than a factor of ten (type C) when scarce or no data were available.

\subsubsection{Vertical mixing}

We describe vertical mixing in the classical way (see, e.g., \citealt{Bauer1973}, \citealt{Yung1999}, \citealt{Catling2017}), whereby the transport flux of a species, $i$, is governed by eddy and molecular diffusion according to
\begin{equation}
\phi_i = - K_{zz} n \frac{\partial f_i}{\partial z} - D_i n \Big( \frac{\partial f_i}{\partial z} + \frac{f_i}{H_i} - \frac{f_i}{H_0} + \frac{\alpha_i}{T} \frac{d T}{ d z} f_i \Big), \label{eq:flux}
\end{equation}
where $K_{zz}$ is the eddy diffusion coefficient, $D_i$ is the coefficient of molecular diffusion of species $i$, $H_i$ is the scale height of species $i$, $H_0$ is the mean scale height of the atmosphere, and $\alpha_i$ is the thermal diffusion factor of species $i$. Eddy diffusion is an empirical formalism to account for the various processes of advection and turbulent mixing in the vertical direction. For the radiative part of the atmosphere, we adopted the recommendation by \cite{Moses2022}, which provides a reasonable fit to $K_{zz}$ values derived for hot Jupiters from GCMs \citep{Parmentier2013,Agundez2014b} and inferred for Solar System atmospheres \citep{Yung1999,Zhang2018},
\begin{equation}
K_{zz} = 5 \times 10^8 p^{-0.5} \bigg( \frac{H_1}{620} \bigg) \bigg( \frac{T_{\rm eff}}{1450} \bigg)^4,
\end{equation}
where $K_{zz}$ is in units of square centimeter per second, $p$ is the pressure in units of bar, $H_1$ is the atmosphere scale height at 1 mbar in units of kilometer, and $T_{\rm eff}$ is the planetary effective temperature in units of degrees Kelvin, with an upper limit of 10$^{11}$ cm$^2$ s$^{-1}$ for the upper layers. For the convective part of the atmosphere, $K_{zz}$ was estimated using free convection \citep{Ackerman2001} as
\begin{equation}
K_{zz} = \frac{H}{3} \bigg(\frac{k_B\,\sigma_{\rm SB}\,T_{\rm int}^4}{m^2\,n\,c_P}\bigg)^{1/3},
\end{equation}
where $k_B$ and $\sigma_{\rm SB}$ are the Boltzmann and Stefan-Boltzmann constants, $T_{\rm int}$ is the internal temperature of the planet in degrees Kelvin, $m$ is the mean atmospheric mass particle in grams, and the quantities $H$ (atmospheric scale height in centimeter), $n$ (volume density of particles per cubic centimeter), and $c_P$ (atmospheric heat capacity at constant pressure in units of ergs per degree Kelvin per gram) were evaluated at the radiative-convective transition layer. The coefficient of molecular diffusion, $D_i$, was evaluated from the kinetic theory of gases \citep{Reid1988} and the factor of thermal diffusion, $\alpha_i$, was set to $-0.25$ for the light species H, H$_2$, and He \citep{Bauer1973}, and to 0 for the rest of species.

Since the atmosphere is discretized into a finite number of layers, the transport fluxes of species $i$ at the lower and upper boundaries of layer $j$, $\phi_i^{j+1/2}$ and $\phi_i^{j-1/2}$, respectively, to be entered in Eq.\,(\ref{eq:continuity-discrete}) are given by
\begin{eqnarray}
\phi_i^{j \pm 1/2} = - K_{zz}^{j \pm 1/2} n^{j \pm 1/2} \frac{\partial f_i}{\partial z} \Big|_{j \pm 1/2} - D_i^{j \pm 1/2} n^{j \pm 1/2} \Bigg[ \frac{\partial f_i}{\partial z} \Big|_{j \pm 1/2} \nonumber \\
+ \bigg( \frac{f_i^{j \pm 1/2}}{H_i^{j \pm 1/2}} - \frac{f_i^{j \pm 1/2}}{H_0^{j \pm 1/2}} + \frac{\alpha_i}{T^{j \pm 1/2}} \frac{d T}{d z} \Big|_{j \pm 1/2} f_i^{j \pm 1/2} \bigg) \Bigg], \label{eq:flux-discrete}
\end{eqnarray}
where the quantities evaluated at the $j$+1/2 and $j$$-$1/2 boundaries are approximated as the arithmetic mean of the values at the $j$ and $j$+1 layers and at the $j$$-$1 and $j$ layers, respectively. We assumed that there is neither gain nor loss of material in the atmosphere, and thus the transport fluxes at the bottom and top boundaries of the atmosphere were set to zero.

\subsection{Chemical equilibrium module}

This module, which is based on the algorithm implemented in the NASA/CEA (Chemical Equilibrium with Applications) program \citep{Gordon1994}, has been used previously to calculate the chemical equilibrium composition of hot-Jupiter atmospheres \citep{Agundez2012,Agundez2014b,Al-Refaie2021,Al-Refaie2024} and AGB atmospheres \citep{Agundez2020}. This piece of code, which is publicly available (see Sect.\ref{sec:data}), computes the chemical equilibrium composition of a gas for a given set of three input parameters: temperature, pressure, and elemental composition. The algorithm is based on the minimization of the Gibbs free energy of a mixture of $N_S$ gaseous species composed of up to $N_E$ elements, which is given by
\begin{equation}
g = \sum_{i=1}^{N_S} \mu_i n_i,
\end{equation}
subject to the condition of mass conservation
\begin{equation}
\sum_{i=1}^{N_S} a_{li} n_i - b_l^0 = 0, \quad (l=1,N_E),
\end{equation}
where $n_i$ is the number of moles of species $i$ per unit mass of mixture, $a_{li}$ is the number of atoms of element $l$ in species $i$, $b_l^0$ is the number of moles of atoms of element $l$ per unit mass of mixture, and the chemical potential for a gaseous species $i$ can be written as
\begin{equation}
\mu_i = \mu_i^o + RT \ln \Big( \frac{n_i}{n} \Big) + RT \ln p,
\end{equation}
where $R$ is the ideal gas constant, $T$ the temperature, $p$ the total pressure, $n$ the total number of moles per unit mass of mixture (and thus $n_i$/$n$ is the mixing ratio or mole fraction of species $i$), and $\mu_i^o$ is the standard-state chemical potential of species $i$, which can be expressed as
\begin{equation}
\mu_i^o = H_i^o(T) - T S_i^o(T),
\end{equation}
where $H_i^o(T)$ and $S_i^o(T)$ are the standard-state enthalpy and entropy, respectively, of species $i$, and standard-state refers to a standard pressure of 1 bar. The thermochemical properties $H_i^o(T)$ and $S_i^o(T)$ are usually parameterized as a function of temperature in the form of NASA polynomial coefficients \citep{McBride2002}. We collected thermochemical data in the form of NASA polynomials for the 164 gaseous species considered mainly from two compilations, the NASA/CEA database\footnote{\tiny \url{https://www1.grc.nasa.gov/research-and-engineering/ceaweb/}} \citep{McBride2002} and the Third Millenium Thermochemical Database\footnote{\tiny \url{https://burcat.technion.ac.il/}} \citep{Goos2018}. The references for the thermochemical data of each species are given in Table\,\ref{tab:species}.

\subsection{Radiative-convective module}

In a 1D planet atmosphere where heat transport is radiatively dominated, the evolution of the vertical distribution of temperature toward radiative equilibrium is given by the equation of thermal energy (e.g., \citealt{Marley2015})
\begin{equation}
\frac{dT}{dt} = \frac{g}{c_P} \frac{dF_{\rm net}}{dp}, \label{eq:dt}
\end{equation}
where $T$, $p$, $g$, and $t$ are temperature, pressure, gravity, and time, respectively, $c_P$ is the local atmospheric heat capacity at constant pressure, and $F_{\rm net}$ is the bolometric net radiative flux (with cgs units of erg s$^{-1}$ cm$^{-2}$), which is given by the difference between the upward (“+”) and downward (“$-$”) radiative fluxes,
\begin{equation}
F_{\rm net} = \int_0^\infty (F_\nu^+ - F_\nu^-)\, d\nu. \label{eq:fnet}
\end{equation}
In principle one could integrate Eq.\,(\ref{eq:dt}) using a uniform time step until radiative equilibrium ($dT/dt=0$) is attained. However, for computational efficiency and numerical stability reasons it is more convenient to use a different time step for each layer according to its own radiative timescale. Here we used the numerical time stepping technique described in \cite{Malik2017} and implemented in the HELIOS code\footnote{\tiny \url{https://github.com/exoclime/HELIOS}}, in which the correction term for the temperature of each layer, $j$, is evaluated as
\begin{equation}
\Delta T_j = \frac{f_{{\rm pre},j} \, p_j}{| \Delta F_{{\rm net},j} |^{0.9}} \frac{\Delta F_{{\rm net},j}}{\Delta p},
\end{equation}
where $f_{{\rm pre},j}$ is a prefactor that takes a value of unity in the first time step and is later on refined down or up depending on whether or not convergence toward radiative equilibrium progresses adequately. Following \cite{Malik2017}, we considered that radiative equilibrium had been attained in each layer, $j$, when the condition
\begin{equation}
\frac{| \Delta F_{{\rm net},j}|}{\sigma_{\rm SB} T_j^4} < 10^{-7}
\end{equation}
was fulfilled, where $\sigma_{\rm SB}$ is the Stefan-Boltzmann constant. To account for the possibility that heat transport may be convectively rather than radiatively dominated in the bottom atmospheric regions, we checked whether the condition
\begin{equation}
T_j > T_{j+1} \bigg( \frac{p_j}{p_{j+1}} \bigg)^{\nabla_{\rm ad}} \label{eq:adiabat}
\end{equation}
held for two adjacent layers, $j$ and $j$+1, and if so, imposed that $T_j$ should be equal to the right-hand side of Eq.\,(\ref{eq:adiabat}). The dry adiabat, $\nabla_{\rm ad}$, is equal to $k_B$/($m\,c_P$), where $k_B$ is the Boltzmann constant and $m$ the mean atmospheric mass particle (e.g., \citealt{Marley2015}).

\begin{figure*}
\centering
\includegraphics[angle=0,width=\textwidth]{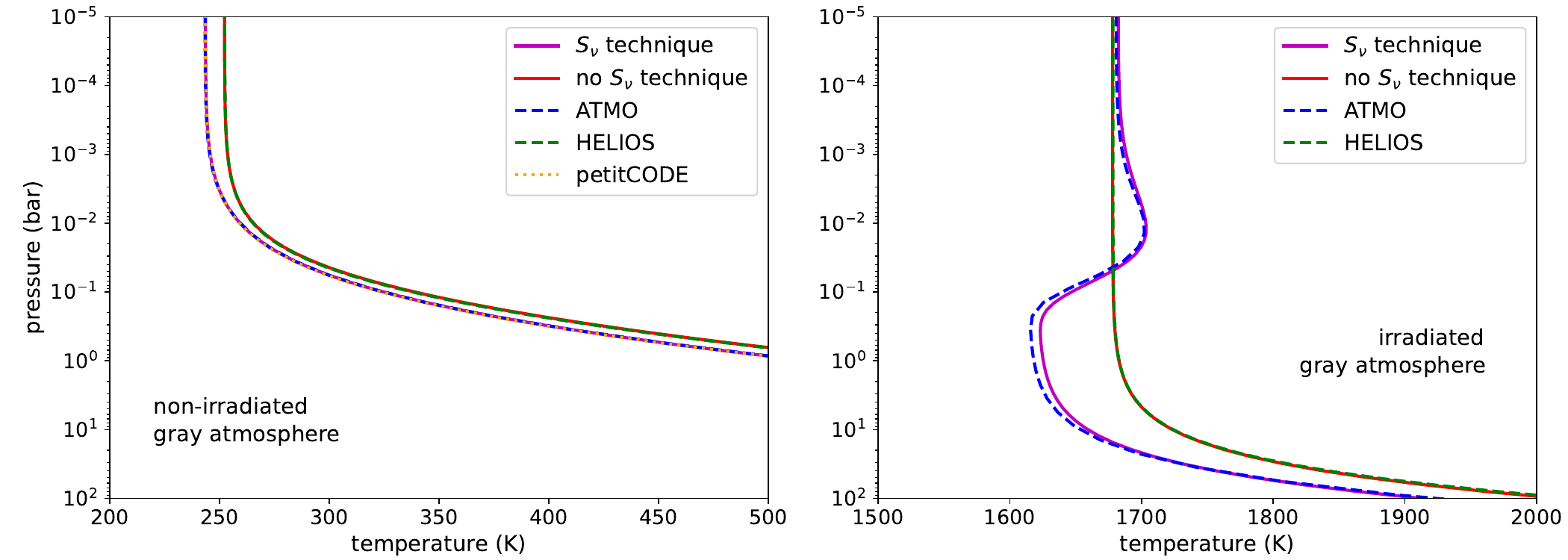}
\caption{Pressure-temperature profiles calculated with our radiative-convective module with and without the source function ($S_\nu$) technique (see text) are compared to those obtained with other codes in the literature: ATMO \citep{Tremblin2015}, HELIOS \citep{Malik2017,Malik2019}, and petitCODE \citep{Molliere2015,Molliere2017}. We consider two cases with parameters typical of HD\,209458b. The first one is a non-irradiated gray atmosphere (left panel) with an absorption coefficient of 0.01 cm$^2$ g$^{-1}$, an internal planet temperature of 300 K, a mean atmosphere molecular weight of 2.3 amu, a planet mass of 219.3 M$_\oplus$, and a planet radius of 15.468 R$_\oplus$. The second case is an irradiated gray atmosphere (right panel) with the same parameters as the previous non-irradiated case and a zenith angle of 60$^\circ$, a surface albedo of 0, a heat redistribution factor of 1 (i.e., no day-night heat redistribution), and a star with a radius of 1.118 R$_{\odot}$ radiating as a blackbody at a temperature of 6000 K and located at a distance of 0.047 AU from the planet. In the non-irradiated case, our “$S_\nu$” result cannot be visually distinguished from those by ATMO and petitCODE, while the same happens between our “no $S_\nu$” result and that from HELIOS. In the irradiated case, our “no $S_\nu$” result overlaps with the HELIOS one.} \label{fig:temperature_benchmark}
\end{figure*}

The calculation of the upwelling and downwelling radiative fluxes $F_\nu^+$ and $F_\nu^-$ in Eq.~(\ref{eq:fnet}) requires one to solve the radiative transfer equation in a vertically inhomogeneous scattering, absorbing, and emitting plane-parallel medium, which reads
\begin{eqnarray}
\mu \frac{dI_\nu(\tau_\nu, \mu, \phi)}{d\tau_\nu} = I_\nu(\tau_\nu, \mu, \phi) - S_\nu(\tau_\nu, \mu, \phi) - \frac{w_{0,\nu}}{4\pi} \cdot \nonumber \\
\cdot \int_0^{2\pi} \int_{-1}^{+1} P_\nu(\mu,\mu'\phi,\phi') I_\nu(\tau_\nu, \mu', \phi') d\mu' d\phi', \label{eq:radiative_transfer}
\end{eqnarray}
where $\nu$ is the frequency, $\mu$ is the cosine of the zenith angle, $\phi$ is the azimuth angle, $\tau_\nu$ is the optical depth (increasing from top to bottom), $I_\nu$ is the specific intensity, $S_\nu$ is the source function, $w_{0,\nu}$ is the single scattering albedo, and $P_\nu$ is the scattering phase function. Here we used the two-stream approximation described in \cite{Toon1989}\footnote{We note that there are a couple of typographical errors in \cite{Toon1989}, which are important to be aware of for a correct coding of the method. In their Eq.\,(40) the running index $l$\,=\,2$n$\,$-$\,1 should read $l$\,=\,2$n$\,+\,1, while in their Eq.\,(42), the expression for $E_l$ should read $E_l$\,=\,[$C_{n+1}^+$(0)\,$-$\,$C_n^+$($\tau_n$)]$e_{2n+1}$ + [$C_n^-$($\tau_n$)\,$-$\,$C_{n+1}^-$(0)]$e_{4n+1}$. These errors were also noted in the code CLIMA developed by J. Kasting ({\tiny \url{https://github.com/VirtualPlanetaryLaboratory/atmos}}, see also p. 451 of \citealt{Catling2017}).}, which was originally based on the two-stream method introduced to treat solar energy deposition into Earth atmosphere \citep{Meador1980} and was found to be numerically stable and computationally efficient. In the two-stream approximation only two directions, upward (“+”) and downward (“$-$”) were considered, so that the method allowed us to compute the monochromatic upward and downward fluxes, $F_\nu^+$ and $F_\nu^-$, respectively, at each interface between atmospheric layers. Moreover, following \cite{Toon1989} we carried out the radiative transfer separately for the stellar incident radiation and for the outgoing planetary thermal emission. Concretely, the penetration of stellar radiation is described using the quadrature scheme \citep{Meador1980,Toon1989} with delta scalings \citep{Joseph1976}. On the other hand, the propagation of thermal emission is described using the hemispheric mean scheme, where we adopt the source function technique, which is supposed to improve the accuracy in the limit of no scattering \citep{Toon1989}.

The radiative-convective module has been benchmarked against several similar codes such as ATMO \citep{Tremblin2015}, HELIOS \citep{Malik2017,Malik2019}, and petitCODE \citep{Molliere2015,Molliere2017}. We note that ATMO and petitCODE have been in turn benchmarked in \cite{Baudino2017}. To that purpose we consider a gray non-irradiated and irradiated atmosphere with a uniform absorption coefficient of 0.01 cm$^2$ g$^{-1}$. The results of the benchmark exercise are shown in Fig.\,\ref{fig:temperature_benchmark}, while the model parameters are given in the foot of this figure. It is seen that in the non-irradiated case, the calculated pressure-temperature profile agrees perfectly with those from ATMO and petitCODE, while that from HELIOS agrees perfectly with ours when the source function technique is not used. In the case of the irradiated atmosphere, our result adopting the source function technique agrees reasonable well with that from ATMO, while that from HELIOS again agrees perfectly with ours when the source function technique is skipped.

Of course, real planet atmospheres are not gray, and thus wavelength-dependent optical depths are needed to solve the radiative transfer equation, Eq.\,(\ref{eq:radiative_transfer}). We computed in advance the wavelength-dependent absorption cross section of each species over a grid of fixed pressures and temperatures using line lists such as ExoMol \citep{Tennyson2024} and HITRAN \citep{Gordon2022} (the complete set of 56 species providing atmospheric opacity that are included for the computation of temperature is given in Table\,\ref{tab:species} and the corresponding reference is given in the column labeled “IR linelist”) and store them in the form of $k$ tables \citep{Lacis1991,Chubb2021}. Our pressure grid covers from 10$^{-6}$ to 10$^3$ bar, with 10 values equally spaced in the decimal logarithm of pressure, while the temperature grid covers from 100 to 3000 K, with 20 values spaced by 50 K from 100 K to 300 K, by 100 K up to 1000 K, by 200 K up to 1800 K, and by 300 K up to 3000 K. The absorption cross section of each species at a given pressure and temperature was computed using a fine wavenumber grid with a uniform step of 10$^{-3}$ cm$^{-1}$ over the 10-10$^5$ cm$^{-1}$ wavenumber range. To reduce the computation time, we used the algorithm of \cite{Amundsen2014}, where the line wing cutoff and whether the line is unimportant and can be skipped are evaluated on the fly for each line. This scheme was not adopted for H$_2$O and CO, where we fixed the cutoff to 25 cm$^{-1}$ \citep{Clough1989} and 100 cm$^{-1}$ \citep{Amundsen2014}, respectively. We then divided the full spectral range into a coarser grid of 456 spectral bins, with the spacings given by the spectral resolution, $\lambda$/$\Delta \lambda$, set to 50, and computed the $k$ distribution $\log k(g)$, where $g$ is the cumulative frequency distribution of the absorption coefficient, $k$ \citep{Lacis1991}, for each spectral bin, and approximated the $k$ distribution using a set of 36 Gauss-Legendre quadrature points (20, 10, and 6 points covering the ranges $g$\,=\,0-0.99, 0.99-0.999, and 0.999-1, respectively), which are conveniently stored. Once pre-calculated $k$ tables were available for each species, we used resorting and rebinning to combine $k$ coefficients of different species \citep{Amundsen2017} and calculated the transmission $T_\nu$, where $T_\nu$\,=\,$\exp(-\tau_\nu)$, in each spectral bin (centered at a frequency, $\nu$) and atmospheric layer.

\section{Models} \label{sec:models}

\begin{table*}
\small
\caption{Stellar and planetary model parameters.}
\label{tab:parameters}
\centering
\begin{tabular}{l@{\hspace{0.38cm}}c@{\hspace{0.38cm}}c@{\hspace{0.38cm}}c@{\hspace{0.38cm}}c@{\hspace{0.38cm}}c@{\hspace{0.38cm}}c@{\hspace{0.38cm}}c@{\hspace{0.38cm}}l}
\hline \hline
Planet             & $R_\star$ (R$_\odot$) & $T_\star$ (K)      & $a$ (AU)      & $M_p$ (M$_\oplus$) & $R_p$ (R$_\oplus$) & $f_i$ & $T_{\rm int}$ (K) & Atmosphere                                               \\
                   &                       & star spectrum      &               &                    &                    &       &                   & chemical composition                                     \\\hline
\multicolumn{8}{c}{Real exoplanet atmospheres} \\
\hline
WASP-33b           & 1.444\,$^a$           & 7400\,$^{a,e}$     & 0.02565\,$^j$ & 667.4\,$^m$        & 17.97\,$^p$        & 1     & 300               & H:He:C:N:O:S:P:Si:Ti solar\,$\times$\,10\,$^{q,r}$                         \\
HD\,209458b        & 1.203\,$^b$           & 6092\,$^{b,f}$     & 0.04747\,$^k$ & 226.9\,$^k$        & 15.468\,$^k$       & 1/2   & 300               & H:He:C:N:O:S:P:Si:Ti solar\,$^{q,s}$                         \\
HD\,189733b        & 0.805\,$^b$           & 4875\,$^{b,g}$     & 0.031\,$^l$   & 371.9\,$^l$        & 12.902\,$^b$       & 1/2   & 300               & H:He:C:N:O:S:P:Si:Ti solar\,$^{q,t}$                         \\
GJ\,436b           & 0.464\,$^c$           & 3684\,$^{c,h}$     & 0.02887\,$^k$ & 18.55\,$^n$        & 3.645\,$^n$        & 1/2   & 300               & H:He:C:N:O:S:P solar\,$\times$\,100\,$^{q,u}$                   \\
GJ\,1214b          & 0.216\,$^d$           & 3026\,$^{d,i}$     & 0.01411\,$^d$ & 8.17\,$^o$         & 2.742\,$^o$        & 1/2   & 300               & H:He:C:N:O:S:P solar\,$\times$\,1000\,$^{q,v}$                 \\
CO$_2$/N$_2$ hot   & 1.0                   & 4875\,$^{b,g}$     & 0.1           & 1.0                & 1.0                & 1/2   & 50                & CO$_2$:N$_2$ 1:1                                         \\
H$_2$O/CO$_2$ cool & 1.0                   & 4875\,$^{b,g}$     & 1.0           & 1.0                & 1.0                & 1/2   & 50                & H$_2$O:CO$_2$ 1:1                                        \\
volcano hot        & 1.0                   & 4875\,$^{b,g}$     & 0.1           & 1.0                & 1.0                & 1/2   & 50                & H$_2$:H$_2$O:CO:CO$_2$:N$_2$:H$_2$S:SO$_2$ 1:1:1:1:1:1:1 \\
N$_2$/CH$_4$ hot   & 1.0                   & 4875\,$^{b,g}$     & 0.1           & 1.0                & 1.0                & 1/2   & 50                & N$_2$:CH$_4$ 1:1                                         \\
N$_2$/O$_2$ warm   & 1.0                   & 4875\,$^{b,g}$     & 0.3           & 1.0                & 1.0                & 1/2   & 50                & N$_2$:O$_2$ 1:1                                          \\
\hline
\end{tabular}
\tablenotea{\\
Note.-- $R_\star$ and $T_\star$ are the stellar radius and effective temperature, respectively, $a$ is the planet-star distance, $M_p$ and $R_p$ are the planetary mass and radius, respectively, $f_i$ is the factor by which the insolation is reduced to compute the dayside temperature as a result of the redistribution of heat from the dayside to the nightside (it takes a value of 1.0 for no redistribution from the dayside to the nightside and 0.5 for full redistribution between dayside and nightside), and $T_{\rm int}$ is the Stefan-Boltzmann temperature associated with the internal heat flux of the planet.\\
$^a$\,\cite{CollierCameron2010}. $^b$\,\cite{Boyajian2015}. $^c$\,\cite{Torres2007}. $^d$\,\cite{Harpsoe2013}.\\
$^e$\,0.001-0.3 $\mu$m: WASP-17 MUSCLES\footref{muscles} spectrum \citep{Behr2023}; 0.3-160 $\mu$m: Castelli-Kurucz\footref{castelli-kurucz} spectrum \citep{Castelli2004} for $T_{\rm eff}$\,=\,7500 K, $\log g$\,=\,4.5, [M/H]\,=\,0.0 scaled to $T_{\rm eff}$\,=\,7400 K; 160-1000 $\mu$m blackbody spectrum.\\
$^f$\,0.00005-0.168 $\mu$m: Sun spectrum (WHI; \citealt{Woods2009}); 0.168-300 $\mu$m: HD\,209458 Kurucz\footref{kurucz} spectrum scaled from $T_{\rm eff}$\,=\,6100 K to 6092 K; 300-1000 $\mu$m blackbody spectrum.\\
$^g$\,0.00051-0.335 $\mu$m: $\epsilon$ Eridani spectrum (provided by I. Ribas, see \citealt{2012VEN}); 0.335-300 $\mu$m: HD\,189733 Kurucz\footref{kurucz} spectrum scaled from $T_{\rm eff}$\,=\,5050 K to 4875 K; 300-1000 $\mu$m blackbody spectrum.\\
$^h$\,0.00055-5.5 $\mu$m: GJ\,436 MUSCLES\footref{muscles} spectrum \citep{France2016,Youngblood2016,Loyd2016}, gap in range 0.16725-0.18065 $\mu$m filled with a uniform brightness of 10$^{-11}$ erg s$^{-1}$ cm$^{-2}$ Hz$^{-1}$ sr$^{-1}$; 5.5-160 $\mu$m: Castelli-Kurucz\footref{castelli-kurucz} spectrum \citep{Castelli2004} for $T_{\rm eff}$\,=\,3750 K, $\log g$\,=\,5.0, [M/H]\,=\,$-$0.5 scaled to $T_{\rm eff}$\,=\,3684 K; 160-1000 $\mu$m blackbody spectrum.\\
$^i$\,0.00055-5.5 $\mu$m: GJ\,1214 MUSCLES\footref{muscles} spectrum \citep{France2016,Youngblood2016,Loyd2016}, gap in range 0.13375-0.18155 $\mu$m filled with a uniform brightness of 10$^{-11}$ erg s$^{-1}$ cm$^{-2}$ Hz$^{-1}$ sr$^{-1}$; 5.5-160 $\mu$m: Castelli-Kurucz\footref{castelli-kurucz} spectrum \citep{Castelli2004} for $T_{\rm eff}$\,=\,3500 K, $\log g$\,=\,5.0, [M/H]\,=\,+0.5 scaled to $T_{\rm eff}$\,=\,3026 K; 160-1000 $\mu$m blackbody spectrum.\\
$^j$\,\cite{Smith2011}. $^k$\,\cite{Southworth2010}. $^l$\,\cite{Paredes2021}. $^m$\,\cite{Lehmann2015}. $^n$\,\cite{Melo2024}. $^o$\,\cite{Cloutier2021}. $^p$\,\cite{Hardy2015}. $^q$\,\cite{Asplund2009}. $^r$\,\cite{Cont2022}. $^s$\,\cite{Xue2024}. $^t$\,\cite{Finnerty2024}. $^u$\,\cite{Mukherjee2025b}. $^v$\,\cite{Schlawin2024}.
}
\end{table*}

We investigated the mutual influence between disequilibrium composition and temperature in various types of planet atmospheres, with different degrees of insolation, UV irradiation, and elemental composition. We first considered atmospheres typical of the most widely observed types of exoplanets: ultrahot Jupiter, hot Jupiter, and warm Neptune. Ultrahot Jupiters are extremely irradiated gas giant planets with equilibrium temperatures in excess of 2000 K \citep{Parmentier2018}. We took as an example of an ultrahot Jupiter WASP-33b, whose atmosphere has been widely characterized through observations. The equilibrium temperature has been measured to be 2784\,$\pm$\,46 K \citep{Chakrabarty2019}, and there is evidence of a dayside temperature inversion \citep{Haynes2015,vanSluijs2023} and of several species, such as Fe, TiO, CO, and OH \citep{Nugroho2020,Nugroho2021,Cont2021,Yan2022}, that are indicative of very high dayside temperatures. In the category of hot Jupiters we consider the renowned planets HD\,209458b and HD\,189733b, which have equilibrium temperatures of 1740 K and 1429 K, respectively \citep{Dang2025}. Current evidence indicates that none of them possess a stratosphere \citep{Schwarz2015,Line2016}, and recent JWST observations have firmly identified H$_2$O and CO$_2$, but no CH$_4$, in their atmospheres \citep{Xue2024,Fu2024}. The chemical processes at work in their atmospheres have been modeled, with it being found that the chemical composition of HD\,209458b is closer to chemical equilibrium than that of HD\,189733b, which is more affected by photochemistry \citep{Moses2011,2012VEN,Agundez2014b}. We also considered the two cooler sub-Jupiter planets GJ\,436b and GJ\,1214b, which have equilibrium temperatures of 583 K and 555 K, respectively \citep{Melo2024,Charbonneau2009}. According to observations, the atmosphere of GJ\,436b seems to be rich in CO and poor in CH$_4$ \citep{Line2014}. Chemical models constructed for these two planets have shown the great sensitivity of the chemical composition to the adopted metallicity \citep{Moses2013,Agundez2014a,Hu2015,Miguel2015,Miller-Ricci-Kempton2012,Hu2014}, and there is observational evidence of an enhanced metallicity in these two planets \citep{Kempton2023,Schlawin2024,Mukherjee2025b}.

The stellar and planetary parameters adopted for WASP-33b, HD\,209458b, HD\,189733b, GJ\,436b, and GJ\,12143b, along with the corresponding references, are listed in Table\,\ref{tab:parameters}. For WASP-33b, HD\,209458b, and HD\,189733b we included all elements and adopted a solar metallicity \citep{Cont2022,Xue2024,Finnerty2024}. To explore the effects of enhanced metallicities in GJ\,436b and GJ\,1214b, we adopted 100 and 1000 times the solar metallicity, respectively, and neglected the refractory elements Si and Ti, which are expected to be in the form of solids at the atmospheric temperatures of these two planets. The stellar spectra adopted were built from different sources. For the UV-visible part, we made extensive use of the MUSCLES\footnote{\tiny \url{https://archive.stsci.edu/prepds/muscles/} \label{muscles}} database, where spectra for GJ\,436 and GJ\,1214 are available, while in the case of WASP-33 we adopted as proxy the spectrum of WASP-17, which has the closest effective temperature among the stars in the MUSCLES sample \citep{Behr2023}. As spectra of HD\,209458 and HD\,189733 we used that of the Sun (WHI; \citealt{Woods2009}) and that of $\epsilon$\,Eridani, respectively, as proxies based on their similar stellar properties. For the infrared part, we used the Kurucz\footnote{\tiny \url{http://kurucz.harvard.edu/stars.html} \label{kurucz}} and Castelli-Kurucz\footnote{\tiny \url{https://wwwuser.oats.inaf.it/fiorella.castelli/grids.html} \label{castelli-kurucz}} databases, while at wavelengths longer than 160-300 $\mu$m we assumed a blackbody spectral shape. More details are given in the foot of Table\,\ref{tab:parameters}.

We also considered secondary atmospheres that may result from exchange processes between the atmosphere and the outer space, the planetary surface, or the planet interior. A nice example of the impact of disequilibrium composition on the temperature in a secondary atmosphere is found on Earth, where the photochemical origin of ozone gives raise to a temperature inversion. The secondary atmospheres considered here are inspired by the telluric planets of our Solar System, but are also based on plausibility for exoplanet atmospheres not yet characterized through observations. Among the variety of plausible secondary-atmosphere compositions \citep{Leconte2015}, we consider atmospheres dominated by H$_2$O and/or CO$_2$, by volcanic outgassing, with the typical molecules outgassed by volcanoes like CO$_2$, H$_2$O, SO$_2$, H$_2$S, H$_2$, and CO \citep{Lee2018}, reducing atmospheres composed of CH$_4$, and oxidizing atmospheres dominated by O$_2$. For simplicity, we assume that the molecules present initially have equal mixing ratios. The planetary parameters adopted are those of a terrestrial-like planet, with the radius and mass of the Earth and illuminated by a star with intense UV emission (such as $\epsilon$\,Eridani). We consider the planet to be located at 0.1, 0.3, or 1.0 AU from the star, so that we cover the cases of hot, warm, and cool atmospheres. Since we are interested in the general behavior of how disequilibrium chemistry affects the temperature, we concentrate on five particular secondary atmospheres, whose parameters are given in Table\,\ref{tab:parameters}. Other cases with a different chemical composition or degree of insolation are not shown here but their behavior can be rationalized into one of the five cases given in Table\,\ref{tab:parameters}, as is discussed in Sect.\,\ref{sec:discussion}.

For simplicity, in all planets modeled the pressure at the bottom level of the atmosphere was set to 100 bar, and we adopted a Bond albedo of 0 and a emissivity of 1 for the surface (or the atmosphere bottom). Since we were interested in modeling the dayside, we considered a zenith stellar illumination angle of 48$^\circ$, in both the computation of the temperature and the photochemistry (see, e.g., \citealt{2021TSA}). Observed phase curves of hot and ultrahot Jupiters \citep{Zhang2018c,Dang2025} indicate that there is an important redistribution of heat in hot Jupiters such as HD\,209458b and HD\,189733b, while the day-to-night heat flow is significantly smaller for ultrahot Jupiters such as WASP-33b. This is in line with theoretical considerations that point to a more efficient heat redistribution for cooler planets \citep{Koll2022}. We therefore assumed in the computation of the dayside temperature that heat is efficiently transported from the dayside to the nightside and thus insolation is reduced by a factor of two ($f_i$\,=\,1/2, which corresponds to $f$\,=\,1/4 in the nomenclature of \citealt{Hansen2008} and to $\varepsilon$\,=\,1 in the nomenclature of \citealt{Cowan2011}), at the exception of WASP-33b, in which case we assume no redistribution of heat from the dayside to the nightside ($f_i$\,=\,1, $f$\,=\,2/3 in the nomenclature of \citealt{Hansen2008}, and $\varepsilon$\,=\,0 in the nomenclature of \citealt{Cowan2011}). There is evidence from the observed inflated radii of hot Jupiters that there must be an intense heat flux arising from the planet interior \citep{Ginzburg2015,Thorngren2019,Komacek2022}. Based on this evidence, we adopted a round value of 300 K as internal temperature for WASP-33b, HD\,209458b, HD\,189733b, GJ\,436b, and GJ\,1214b, while a lower round value of 50 K was adopted for the planets with generic secondary atmospheres, in line with the range 20-40 K inferred for Earth, Mars, and Venus \citep{Davies2010,Smrekar2012,Parro2017}.

\begin{figure}
\centering
\includegraphics[angle=0,width=\columnwidth]{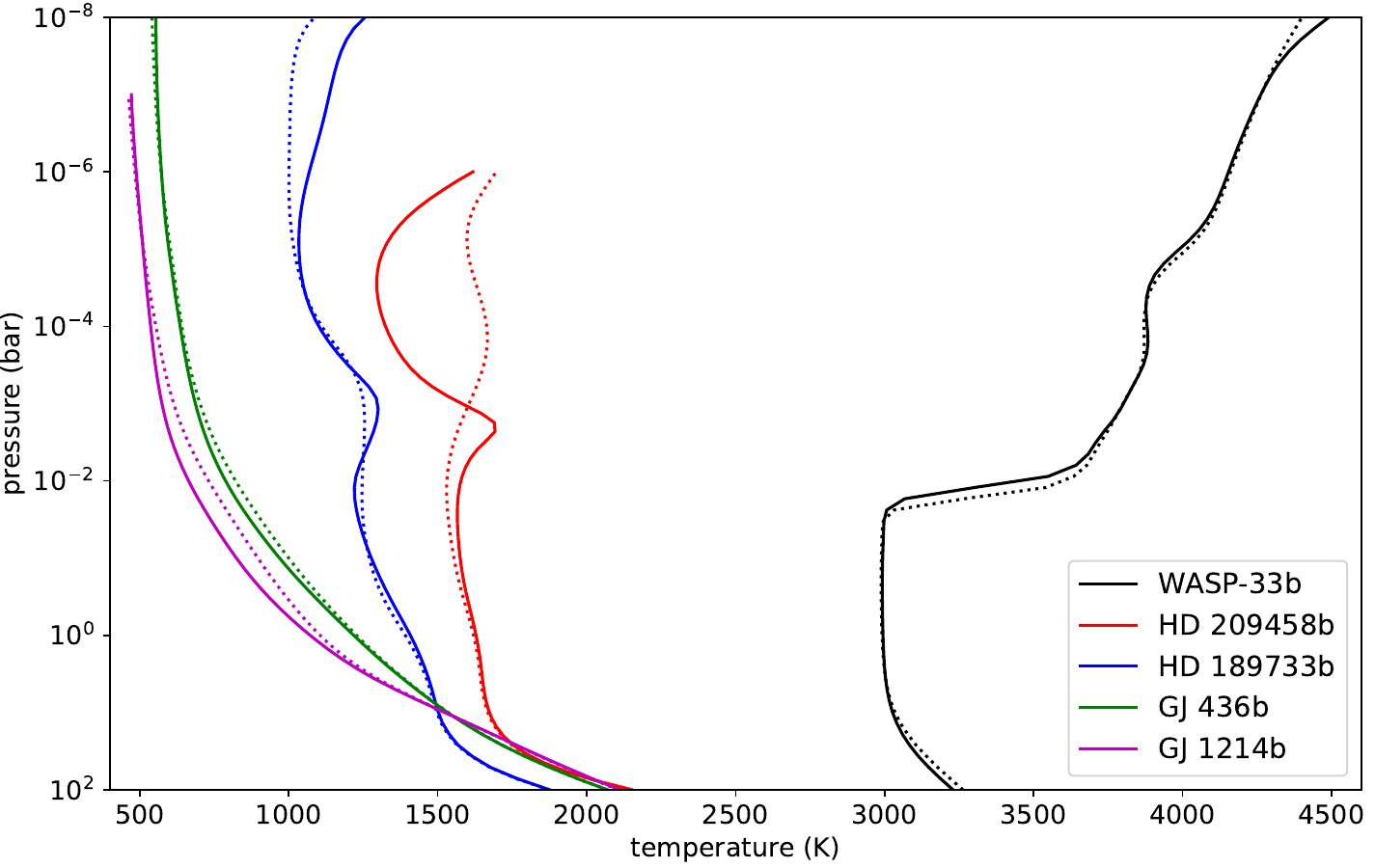}
\caption{Dayside pressure-temperature profiles calculated for the gas giant exoplanets WASP-33b, HD\,209458b, HD\,189733b, GJ\,436b, and GJ\,1214b. The solid lines correspond to the temperature profiles calculated self-consistently with the disequilibrium composition at a time at which a steady state is reached in the atmosphere composition and temperature, while the dotted lines correspond to temperature profiles calculated at the initial time, where the composition is assumed to be given by chemical equilibrium.} \label{fig:temperature_primary}
\end{figure}

\section{Results} \label{sec:results}

\begin{figure*}
\centering
\includegraphics[angle=0,width=\textwidth]{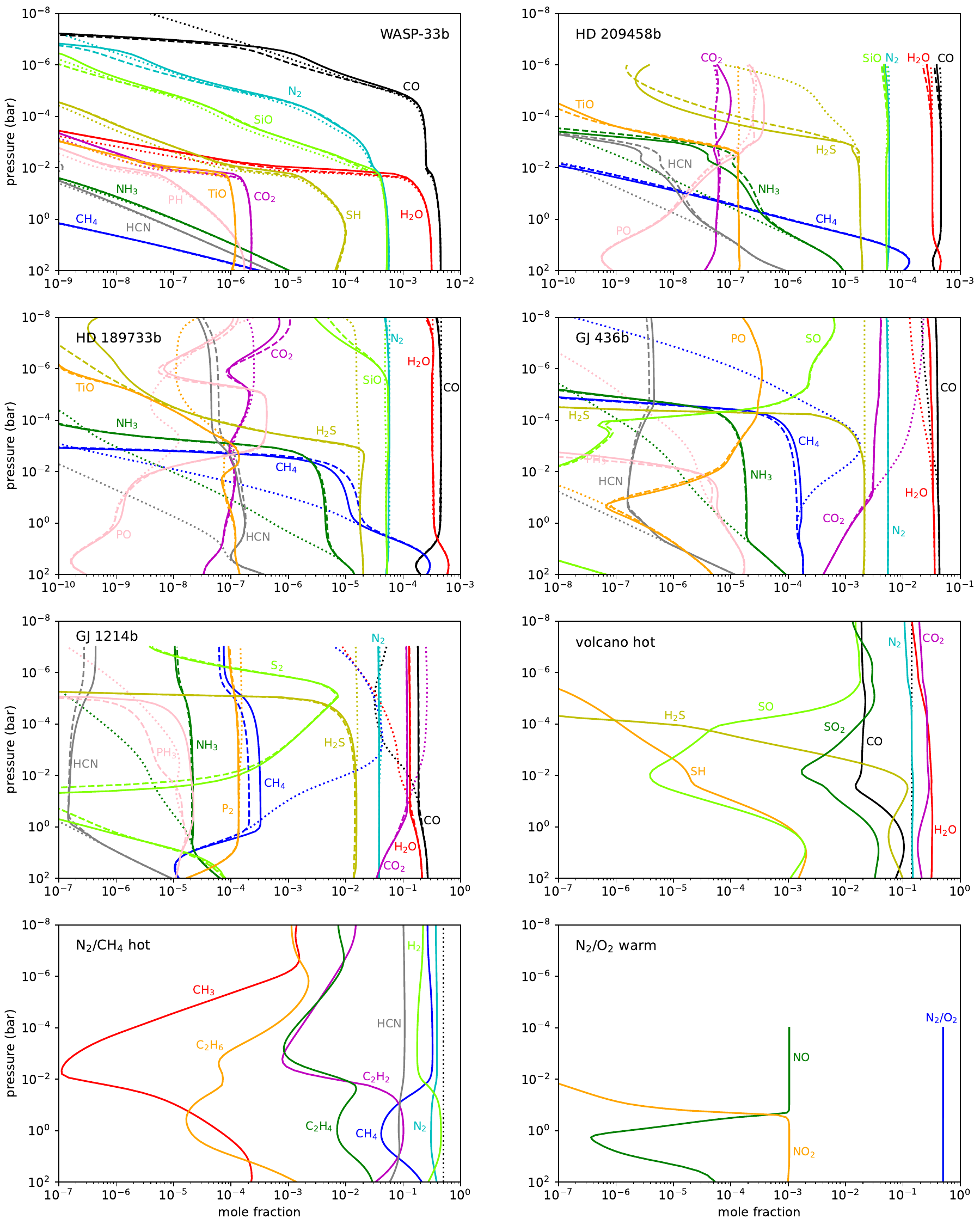}
\caption{Calculated dayside composition in five gas giants and three plausible secondary atmospheres (hot volcano-like, hot reducing N$_2$/CH$_4$, and warm oxidizing N$_2$/O$_2$ atmosphere). Dotted lines correspond to the initial composition (chemical equilibrium for the gas giants and equal uniform abundances for secondary atmospheres), dashed lines to a calculation where the initial temperature does not further evolve (only for the gas giants), and solid lines to a self-consistent calculation where temperature and chemical composition evolve together.} \label{fig:composition}
\end{figure*}

\begin{figure}
\centering
\includegraphics[angle=0,width=\columnwidth]{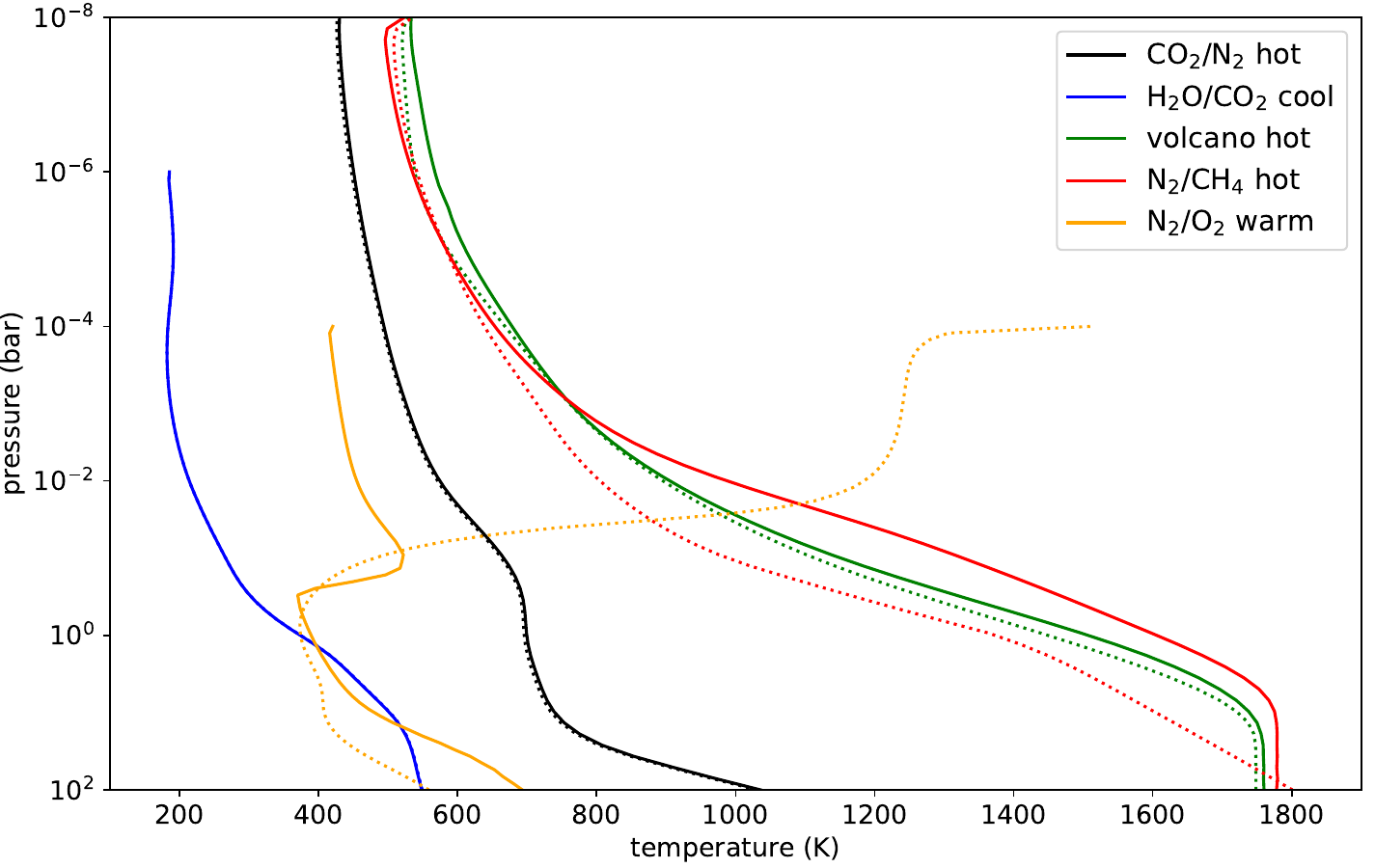}
\caption{Dayside pressure-temperature profiles calculated for five exoplanets with plausible secondary atmospheres (see Table\,\ref{tab:parameters}): a hot atmosphere made of CO$_2$ and N$_2$, a cool atmosphere of H$_2$O and CO$_2$, a hot atmosphere composed of volcanic outgassing, a hot reducing atmosphere of N$_2$ and CH$_4$, and a warm oxidizing atmosphere of N$_2$ and O$_2$. The solid lines correspond to the temperature profiles calculated self-consistently with the disequilibrium composition at a time at which a steady state is reached in the atmosphere composition and temperature, while the dotted lines correspond to temperature profiles calculated at the initial time, where chemistry has not yet modified the initial composition.} \label{fig:temperature_secondary}
\end{figure}

We first examine the influence of disequilibrium composition on the temperature for the five gas giant exoplanets WASP-33b, HD\,209458b, HD\,189733b, GJ\,436b, and GJ\,1214b. In Fig.\,\ref{fig:temperature_primary} we show as solid lines the vertical dayside temperature profile calculated for each of these planets at a time at which a steady state is reached in the chemical composition and temperature. In this calculation, temperature and disequilibrium composition evolve self-consistently. These temperature profiles are compared with those prevailing at the initial time, when the composition is at chemical equilibrium (dotted lines in Fig.\,\ref{fig:temperature_primary}). In general, the differences imprinted by disequilibrium chemistry on the temperature are relatively small. Differences are very minor in the case of the ultrahot Jupiter WASP-33b, significant in the case of the two hot Jupiters HD\,209458b and HD\,189733b, and moderately low (less than 100 K) in the case of the enhanced metallicity warm sub-Jupiters GJ\,436b and GJ\,1214b. In general, if disequilibrium chemistry cause a significant departure of the composition from the chemical equilibrium one, the temperature may vary as a result of the change in the composition. However, in the end the temperature in the radiative part of the atmosphere is essentially controlled by a few atmospheric constituents that are efficient absorbers at infrared and/or visible/UV wavelengths and are abundant enough to dominate the atmospheric opacity in these spectral regions. Therefore, the temperature would only be affected if the abundance of any of these species that dominate the atmospheric opacity is significantly altered due to disequilibrium chemistry.

In the case of the ultrahot Jupiter WASP-33b, the hot and dense bottom atmosphere is essentially molecular (with species like H$_2$, CO, and H$_2$O), while the ultrahot and less dense upper atmospheric layers are mostly atomic. In any case, since temperatures are so high ($>$\,3000 K) anywhere in the dayside, the composition is very close to chemical equilibrium (see top left panel in Fig.\,\ref{fig:composition}) and the temperature is not affected by disequilibrium chemistry. This conclusion should be generalized to any ultrahot Jupiter. The small departures from chemical equilibrium seen in the composition (top left panel in Fig.\,\ref{fig:composition}) are due to vertical mixing for regions below the 10$^{-6}$ bar level and to photochemistry in upper layers.

The hot Jupiter HD\,209458b shows a moderately large impact of disequilibrium chemistry on the temperature. The main difference of the self-consistent temperature profile compared to the chemical equilibrium one consists of a small temperature inversion around 1 mbar and a cooler upper atmosphere, where the temperature drops by about 300 K. There are some species whose disequilibrium abundance depart significantly from the chemical equilibrium one (see top right panel in Fig.\,\ref{fig:composition}), but the only molecule responsible for the temperature modification seen in Fig.\,\ref{fig:temperature_primary} is TiO. Chemical equilibrium predicts a roughly uniform abundance for TiO along the vertical direction, while disequilibrium predicts a sharp abundance decline at pressures below 1 mbar. In our model, the disappearance of TiO in the upper atmosphere is caused by photodissociation, although it must be noted that there are big uncertainties regarding the photochemistry of this species (see Sect.\,\ref{sec:discussion}). In the case of the hot Jupiter HD\,189733b, the self-consistently computed temperature shows some deviations from the chemical equilibrium one, which consist of the appearance of a small temperature inversion at 1 mbar and a warming of the uppermost layers by about 150 K. These two differences are also caused, as in the case of HD\,209458b, by photodestruction of TiO. Concretely, the slight stratosphere at 1 mbar is caused by the abundance decline of TiO in the upper atmosphere (see upper middle left panel in Fig.\,\ref{fig:composition}), while the warming of the upper atmosphere is due to the abundance enhancement of atomic titanium, which is also an important absorber at visible wavelengths \citep{Piskunov1995,Kupka1999,Ryabchikova2015}. 

The warm sub-Jupiter planets GJ\,436b and GJ\,1214b show a similar behavior regarding the effect of disequilibrium composition on the temperature, in spite of the very different metallicities assumed. The main effect consists of a slight cooling of the atmosphere ($<$\,100 K) between 1 and 10$^{-4}$ bar (see Fig.\,\ref{fig:temperature_primary}). For both planets there are important differences in the disequilibrium composition compared to the chemical equilibrium one (see upper middle right and lower middle left panels in Fig.\,\ref{fig:composition}), although the only species that is at the origin of the change in the temperature is CH$_4$, whose abundance is significantly reduced due to the quenching effect caused by vertical mixing. It is worth noting that S- and P-bearing molecules such as H$_2$S, S$_2$, SO, SO$_2$, PH$_3$, P$_2$, and PO are predicted to be present with non-negligible abundances and are thus candidates to be detected. In fact, there is evidence of SO$_2$ in the atmosphere of GJ\,3470b \citep{Beatty2024}, which is a warm sub-Jupiter that resembles GJ\,436b \citep{Venot2014}. The hydrides H$_2$S and PH$_3$ are restricted to the bottom atmosphere because they are readily photodissociated in the upper atmosphere, P$_2$ maintains a high abundance throughout the whole atmosphere following the chemical equilibrium profile, and S$_2$, SO, SO$_2$, and PO are enhanced in the upper layers due to photochemistry. We note, however, that the chemistry of phosphorus is subject to significant uncertainties regarding the thermochemistry of P oxides \citep{Bains2023}, the kinetics of the interconversion between PH$_3$ and phosphorus oxides \citep{Wunderlich2023}, and the photodissociation cross sections of PO and P$_2$, which are not known.

In secondary atmospheres dominated by different initial mixtures of gases, chemical and photochemical processes can lead to important changes in the composition that may affect the atmospheric temperature. In Fig.\,\ref{fig:temperature_secondary} we compare self-consistently calculated temperature profiles with those computed at the initial time, when the composition consists of (vertically uniform) equal mixing ratios for all the species initially present. For the hot atmosphere made of CO$_2$ and N$_2$ and the cool atmosphere of H$_2$O and CO$_2$ the temperature remains unchanged. These two planets have in common that CO$_2$ and/or H$_2$O are dominant atmospheric constituents. These two molecules are very stable, and thus their abundances remain essentially unchanged throughout most of the atmosphere, and very efficient absorbers of infrared light, so that they dominate the radiative balance and thus set the temperature in most of the atmosphere. Even if new species that are also active absorbers of infrared radiation, such as CO, are formed by disequilibrium chemistry, they are less abundant than CO$_2$ and/or H$_2$O and do not compete with them in terms of opacity. In the hot volcano-like atmosphere, which resembles that on Venus \citep{Fegley2014} and could be common among rocky exoplanets \citep{Bello-Arufe2025}, the changes in the temperature are very small (see Fig.\,\ref{fig:temperature_secondary}). This case is similar to the two aforementioned cases in that the atmosphere is dominated, among other species, by CO$_2$ and H$_2$O, and thus, even if disequilibrium chemistry is able to modify the initial abundances and to produce species not initially present, such as SO (which is formed due to photochemistry), the fact that CO$_2$ and H$_2$O remain as dominant species (see lower middle right panel in Fig.\,\ref{fig:composition}) results in a low impact for the temperature.

We now focus on other plausible secondary atmospheres that do not contain CO$_2$ or H$_2$O as dominant constituents, namely a reducing atmosphere made of N$_2$ and CH$_4$, which resembles that of Titan \citep{Dobrijevic2014,Horst2017}, and an oxidizing atmosphere made of N$_2$ and O$_2$, similar to that on Earth, although in both cases we consider a higher degree of insolation that would correspond to a close-in exoplanet. In the case of the reducing atmosphere, the abundance of N$_2$ remains nearly unaltered, although this is not very relevant for the temperature because N$_2$ is not an efficient absorber of light. However, CH$_4$ is heavily processed into hydrocarbons such as CH$_3$, C$_2$H$_2$, C$_2$H$_4$, and C$_2$H$_6$ (see bottom left panel in Fig.\,\ref{fig:composition}). Below the quench level of CH$_4$, which lies around 10$^{-2}$ bar, the processing is mainly driven by thermochemical kinetics, while in upper layers it is the combined action of photochemistry and vertical mixing which determines the chemical composition. Methane is not severely depleted in the top layers because it is brought by vertical mixing from deeper layers, compensating the effect of photodissociation. The change in the abundance profile of CH$_4$ and the appearance of other important absorbers of radiation leads to a warming of the deep atmospheric layers (see Fig.\,\ref{fig:temperature_secondary}). In the case of the oxidizing atmosphere, since N$_2$ and O$_2$ are very transparent at infrared wavelengths, an atmosphere exclusively made of these two gases would show an inverted temperature profile, with temperature rising with increasing altitude and the upper atmospheric layers having rather elevated temperatures (see Fig.\,\ref{fig:temperature_secondary}). In such an atmosphere, our self-consistent calculations indicate that most of N$_2$ and O$_2$ would remain unaltered (see bottom right panel in Fig.\,\ref{fig:composition}), although disequilibrium chemistry would allow one to form some nitrogen oxides in small but sufficient quantities to provide atmospheric opacity at infrared wavelengths and completely modify the temperature structure of the atmosphere (see Fig.\,\ref{fig:temperature_secondary}).

\section{Discussion} \label{sec:discussion}

The main lesson learned from the modeling exercise of the five gas giant atmospheres is that the impact of disequilibrium composition on the dayside temperature is moderately low, with temperature differences which are in general less than 100 K. Even in HD\,189733b, GJ\,436b, and GJ\,1214b, where there are important departures from chemical equilibrium in the composition, the temperature is not greatly affected. As long as the chemical composition is dominated by, apart from H$_2$ and He, active infrared absorbers such as H$_2$O, CO$_2$, CO, and/or CH$_4$, and their abundances are not strongly modified due to the action of thermochemistry, photochemistry, and/or vertical mixing, the vertical temperature profile prevailing in a disequilibrium composition should be similar to that computed under chemical equilibrium. This conclusion is in agreement with previous studies that focused on the warm Neptune GJ\,436b \citep{Agundez2014a} and hot Jupiters \citep{Drummond2016,Mukherjee2023,Mukherjee2024}. The only exception where we found important temperature differences correspond to the presence of TiO in hot Jupiters. This molecule is known to have a large absorption cross section at visible wavelengths \citep{McKemmish2019}, which makes it to cause temperature inversions in hot Jupiter atmospheres \citep{Fortney2008}. Our calculations predict a severe depletion of TiO in the upper atmosphere due to photodissociation, with an important impact on the temperature. However, the photodissociation cross section adopted here for TiO is just an educated guess since it is to the best of our knowledge not known. Moreover, for TiO the threshold for ionization is slightly smaller than for dissociation \citep{Naulin1997,Huang2013}, which means that photoionization probably competes or even dominates over photodissociation. In addition, the chemical reactions involving TiO are poorly known \citep{1993CAM,2008HIG,2013PLA} and the role of condensation complicates even further the picture \citep{Parmentier2013,Roth2024}. There are therefore big uncertainties on the abundance of TiO in the upper atmosphere of hot Jupiters, and thus on the resulting temperature in these layers.

We are not only interested in the study of the potential effects that an atmospheric composition in disequilibrium may have on the temperature but also the other way around; that is, whether the changes induced on the temperature may affect the chemical composition. To this purpose we compare for the five gas giants the composition calculated by the self-consistent model (solid lines in Fig.\,\ref{fig:composition}) with that calculated when the initial temperature profile (calculated at chemical equilibrium) is not allowed to evolve (dashed lines in Fig.\,\ref{fig:composition}), which is the usual assumption in most chemical models of exoplanet atmospheres \citep{Moses2011,Moses2013,2012VEN,Venot2014,Venot2016,Miller-Ricci-Kempton2012,Miguel2014,Miguel2015,Hu2014,Hu2015,Molaverdikhani2019,Hobbs2019}. The differences in the abundances are very small, of a factor of a few at most. This is expected because the differences in the temperature between the initial chemical equilibrium composition and the final steady state one are already small. Even in the case of HD\,209458b, where the largest temperature differences are found, the impact on the composition is small. In the case of GJ\,1214b, some differences are found for the abundances of CH$_4$ and PH$_3$, but they are small and unlikely to leave an imprint on the spectrum of the planet.

We can also extract some lessons from the models of secondary atmospheres studied here. Essentially, in atmospheres dominated by H$_2$O or CO$_2$ (these two molecules are very stable against photochemical and thermochemical processes and control to a large extent the temperature due to their great efficiency in absorbing infrared light), any disequilibrium chemistry taking place is unlikely to have an important effect on the atmospheric temperature structure. On the other hand, in secondary atmospheres where neither H$_2$O nor CO$_2$ are important atmospheric constituents, such as reducing atmospheres dominated by CH$_4$ or oxidizing atmospheres dominated by O$_2$, the processing of the atmospheric composition by disequilibrium chemistry may have important consequences for the atmospheric temperature.

The models investigated here are 1D models that focus on the dayside, but if one expands the horizons beyond the vertical direction there may be room for interesting feedback effects between the chemistry and the temperature. Highly irradiated tidally locked exoplanets are expected to have a strong temperature contrast between the dayside and the nightside. The vigorous circulation between the two sides could produce interesting mutual effects between composition and temperature. From the lessons learned here, we expect that if important changes occur in the abundance of efficient absorbers, such as H$_2$O, CO$_2$, CO, and/or CH$_4$, the temperature can be significantly affected. The distribution of atmospheric constituents as a function of altitude and longitude has been investigated through pseudo-2D models for the hot Jupiters HD\,209458b, HD\,189733b, and WASP-43b \citep{Agundez2014b,Venot2020}. These studies have found that in spite of the marked temperature contrast between dayside and nightside, the strong zonal wind leads to an important homogenization of the chemical composition with longitude. Since the abundances of the main sources of opacity become quite uniform with longitude, we do not expect the temperature to be greatly affected by the different composition between dayside and nightside. In particular, for H$_2$O and CO there are small abundance variations with longitude, while CH$_4$ and CO$_2$ experience some important variations with longitude, although they are present at a lower level of abundance than H$_2$O and CO, and thus play a minor role in setting the atmospheric temperature. In addition, as the gas circulates between the dayside and the nightside, thermal inertia would attenuate any temperature variation due to a longitudinal change of the chemical composition \citep{Iro2005}. The possible presence of gaseous TiO in hot Jupiters may induce important effects on the temperature, as long as its abundance varies with longitude. It is however unclear how this variation would be or even whether TiO is present in hot Jupiter atmospheres, because TiO is expected to condense in the nightside and may not be able to come back to the gas phase in the dayside \citep{Parmentier2013,Roth2024}, and also because TiO is expected to deplete in the dayside due to photodestruction, as discussed in Sect.\,\ref{sec:results}. For exoplanets cooler than hot Jupiters, the longitudinal homogenization of the chemical composition is even more marked \citep{Baeyens2021,Baeyens2022,Moses2022}, so that there would be little contrast between the composition of the dayside and the nightside, and thus little effect on the temperature.

\cite{Drummond2020} have extended the study of the distribution of the atmospheric composition in HD\,209458b and HD\,189733b to 3D using a self-consistent model that includes thermochemical kinetics, radiative transfer, and hydrodynamics. In line with the above discussion, these authors find that the main species causing opacity, such as H$_2$O and CO, are homogeneously distributed across the three dimensions of the atmosphere. Only CH$_4$ show some abundance variations, mainly with latitude, in HD\,209458b, although this molecule is present with an abundance not high enough to dominate the atmospheric opacity. These conclusions are consistent with previous findings by \cite{Cooper2006}.

A case in which chemical differences between the dayside and the nightside could affect the temperature concerns ultrahot Jupiters. In such planets, the very high temperatures present on the dayside cause hydrogen to be mostly in an atomic form, while the lower temperatures of the nightside favor the molecular form. The circulation between dayside and nightside causes hydrogen to switch between atomic and molecular, and the heat associated with this chemical transformation impacts the atmospheric temperature \citep{Bell2018}. This phenomenon has been studied through a pseudo-2D model by \cite{Roth2021}, leading to appreciable changes in the temperature of several hundred degrees Kelvin. A similar effect could a priori be expected for other chemical transformations, such as CO $\leftrightarrow$ CH$_4$. Indeed, in gas giant planets with effective temperatures between 700 K and 1000 K, chemical equilibrium predicts that CO dominates on the dayside, while CH$_4$ would be more abundant on the nightside \citep{Moses2022}. However, as has already been discussed above and shown by \cite{Moses2022}, such a transformation is inhibited by horizontal circulation and CO dominates at both the dayside and the nightside.

The presence of clouds or hazes may have an impact on the atmospheric temperature. Temperate and cool atmospheres are amenable to clouds formed by the condensation of the less volatile species, while strongly UV-irradiated atmospheres may contain photochemically produced hazes. Both clouds and hazes are known to have an important impact on the emission and transmission spectrum of the planet \citep{Pont2008,Pont2013,Parmentier2018,Parmentier2021}, but they may also affect the atmospheric temperature. Indeed, \cite{Lavvas2021} found that photochemical hazes can impact the atmospheric temperature of HD\,189733b by a few hundred degrees Kelvin. Moreover, if clouds or hazes are unevenly distributed across the dayside and nightside, the heat associated with the phase change can also leave an imprint on the atmospheric temperature.

A last point that is worth discussing is the role of UV opacity on the atmospheric temperature. Here, to compute the temperature we solved the radiative transfer over a wide wavelength range, from 100 nm to 1 mm. That is, in addition to the infrared and visible ranges usually considered, where most opacity data come from line lists such as ExoMol \citep{Tennyson2024} and HITRAN \citep{Gordon2022}, we also included the UV region. At these wavelengths line lists are scarce and are restricted to a few diatomic molecules, such as CH \citep{Masseron2014}, OH \citep{Rothman2010}, O$_2$ \citep{Gordon2022}, CN \citep{Syme2020,Syme2021}, SH \citep{Gorman2019}, S$_2$ \citep{Gomez2024}, SO \citep{Brady2024}, PN \citep{Semenov2025}, SiH \citep{Yurchenko2018b}, and SiO \citep{Yurchenko2022}. To take into account UV absorption for the computation of temperature we have thus adopted UV absorption cross sections, which are usually available at room temperature in specific spectral ranges and with variable degrees of spectral resolution. This approach allows us to treat those cases in which absorption at UV wavelengths by some species is important to establish the atmospheric temperature, as occurs on Earth atmosphere where the presence of ozone leads to a temperature inversion, although it does not correctly describe the effect of temperature on the continuum absorption \citep{Venot2018b} or that of temperature and pressure on the intensity and width of absorption lines when these are important \citep{vanDishoeck1987}. In order to evaluate the importance of UV opacity to establish the atmosphere thermal structure we have run models with and without UV opacity when computing the temperature. We find that in general the impact of UV opacity on the temperature is small. For the gas giant primary atmospheres and the secondary atmospheres investigated here, the changes in the temperature when UV opacity is neglected are just a few tens of degrees Kelvin. The only case in which UV opacity is important corresponds to secondary atmospheres whose main atmospheric constituents are transparent at infrared wavelengths, as occurs for the atmosphere composed of N$_2$ and O$_2$. In this case, the absorption of O$_2$ at UV wavelengths becomes crucial to provide atmospheric opacity and establish the atmospheric temperature.

\section{Conclusions} \label{sec:conclusions}

We have developed a code that solves self-consistently the temperature and disequilibrium chemistry in the vertical direction in a planetary atmosphere. We have applied our code to well-known gas giant exoplanets with different degrees of insolation and metallicity (WASP-33b, HD\,209458b, HD\,189733b, GJ\,436b, and GJ\,1214b) and to exoplanets with plausible secondary atmospheres that could potentially be characterized in the future.

In irradiated gas giants with solar or supersolar metallicities, the corrections to the temperature due to disequilibrium chemistry are small, below 100 K, in agreement with previous studies. Although the atmospheric composition of the cooler and more UV-irradiated planets deviates significantly from chemical equilibrium, the impact on the temperature is moderate because the main atmospheric constituents providing opacity (H$_2$O, CO$_2$, CO, and/or CH$_4$) do not experience drastic changes in their abundances due to disequilibrium chemistry. The largest effect on the temperature, of several hundred degrees Kelvin, is found to be caused by TiO in hot Jupiters. However, the extent of this effect is uncertain due to the pending question of whether or not gaseous TiO is present in such atmospheres and the big uncertainties associated with its chemistry.

For plausible secondary atmospheres, the impact of disequilibrium chemistry on the temperature depends on the exact composition. In general, we anticipate that in atmospheres dominated by H$_2$O and/or CO$_2$ the temperature is unlikely to be affected to an important extent, while reducing atmospheres dominated by CH$_4$ and oxidizing atmospheres composed of O$_2$ may see important changes in the temperature as a result of the severe processing of the composition due to disequilibrium chemistry.

\section{Data availability} \label{sec:data}

The full version of Table\,\ref{tab:reactions} is available at the public repository \href{https://zenodo.org/records/15607694}{zenodo}. The code, named PACT (Planetary Atmosphere Chemistry and Temperature), and the data used in this work can be freely accessed at
{{\tiny \url{https://github.com/marcelinoagundez/pact}}. The chemical equilibrium module can also be freely accessed as an independent code named ACE (Atmospheric Chemical Equilibrium) at {\tiny \url{https://github.com/marcelinoagundez/ace}}. Any update to the codes and data will be posted on the above {\tiny \texttt{github}} links. 

\begin{acknowledgements}

We thank E. H\'ebrard, P. Molli{\`e}re, and M. Malik for their help during the validation of the radiative-convective module, and O. Roncero for useful conversations on TiO. We acknowledge funding support from Spanish Ministerio de Ciencia, Innovaci\'on y Universidades through grant PID2023-147545NB-I00 and the computational resources provided by the DRAGO computer cluster managed by SGAI-CSIC, and the Galician Supercomputing Center (CESGA). We thank the anonymous referee for rising important points that helped to improve the manuscript.

\end{acknowledgements}

\begin{appendix}

\onecolumn
\section{Species and reactions tables}

\begin{table}[h!]
\tiny
\caption{Species list.}
\label{tab:species}
\centering
\begin{tabular}{lccclccclccc}
\hline \hline
Species & \multicolumn{3}{c}{Reference} & Species & \multicolumn{3}{c}{Reference} & Species & \multicolumn{3}{c}{Reference}\\
\cline{2-4} \cline{6-8} \cline{10-12}
& \multicolumn{1}{c}{therm} & \multicolumn{1}{c}{IR linelist} & \multicolumn{1}{c}{UV section} & & \multicolumn{1}{c}{therm} & \multicolumn{1}{c}{IR linelist} & \multicolumn{1}{c}{UV section} & & \multicolumn{1}{c}{therm} & \multicolumn{1}{c}{IR linelist} & \multicolumn{1}{c}{UV section} \\
\hline
H                  &  (1) &       (5) & (37,38)    & CO                 &  (1) & (6,16,17) & (39)       & N$_2$O             &  (1) &   (11,24) & (87,39)    \\
H$_2$              &  (1) &   (6,7,8) & (39)       & HCO                &  (1) &        -  & (39,66)    & N$_2$O$_3$         &  (1) &        -  & (88)       \\
He                 &  (1) &       (5) & (37,40)    & H$_2$CO            &  (2) &      (15) & (67,39,45) & NCO                &  (2) &        -  & (55)       \\
Ar                 &  (1) &        -  & -          & CH$_2$OH           &  (2) &        -  & (55)       & CNO                &  (2) &        -  & (53)       \\
C                  &  (1) &       (5) & (37,41)    & CH$_3$O            &  (2) &        -  & (55)       & HNCO               &  (1) &        -  & (39,45)    \\
CH                 &  (2) &  (6,9,10) & (39)       & CH$_3$OH           &  (2) &      (15) & (39,45)    & HOCN               &  (2) &        -  & (53)       \\
CH$_2$             &  (2) &        -  & (39)       & CO$_2$             &  (1) &      (11) & (68,39,45) & HCNO               &  (2) &        -  & (55)       \\
$^1$CH$_2$         &  (2) &        -  & (39)       & HCOOH              &  (2) &      (15) & (39,45)    & CH$_2$NO           &  (2) &        -  & (53)       \\
CH$_3$             &  (2) &        -  & (42)       & CH$_3$O$_2$        &  (2) &        -  & (69)       & CH$_3$NO           &  (2) &        -  & (55)       \\
CH$_4$             &  (1) &   (11,12) & (43)       & CH$_3$OOH          &  (1) &        -  & (69)       & S                  &  (1) &       (5) & (39,89)    \\
C$_2$              &  (2) & (6,13,14) & (39)       & C$_2$O             &  (2) &        -  & (53)       & SH                 &  (2) &    (6,25) & (39)       \\
C$_2$H             &  (2) &        -  & (39)       & HCCO               &  (2) &        -  & (55)       & H$_2$S             &  (1) &      (15) & (90,39)    \\
C$_2$H$_2$         &  (1) &      (15) & (44)       & CH$_2$CO           &  (1) &        -  & (70)       & S$_2$              &  (1) &      (26) & (39)       \\
C$_2$H$_3$         &  (2) &        -  & (39)       & CH$_3$CO           &  (2) &        -  & (71)       & HS$_2$             &  (2) &        -  & (53)       \\
C$_2$H$_4$         &  (1) &      (15) & (39,45)    & CH$_3$CHO          &  (1) &        -  & (39,45)    & H$_2$S$_2$         &  (4) &        -  & (55)       \\
C$_2$H$_5$         &  (2) &        -  & (39)       & CH$_3$CHOH         &  (2) &        -  & (55)       & CS                 &  (1) &    (6,27) & (39)       \\
C$_2$H$_6$         &  (1) &      (15) & (39,45)    & C$_2$H$_5$O        &  (2) &        -  & (55)       & HCS                &  (2) &        -  & (55)       \\
C$_3$              &  (2) &        -  & (39)       & C$_2$H$_4$OH       &  (2) &        -  & (72)       & H$_2$CS            &  (2) &        -  & (39)       \\
C$_3$H             &  (2) &        -  & (39)       & CH$_3$OCH$_2$      &  (2) &        -  & (73)       & CH$_3$S            &  (2) &        -  & (55)       \\
C$_3$H$_2$         &  (2) &        -  & (39)       & C$_2$H$_5$OH       &  (1) &        -  & (39)       & CH$_3$SH           &  (2) &        -  & (39,45)    \\
C$_3$H$_3$         &  (2) &        -  & (39)       & CH$_3$OCH$_3$      &  (2) &        -  & (39)       & CS$_2$             &  (1) &      (15) & (39)       \\
CH$_3$CCH          &  (1) &        -  & (46)       & N                  &  (1) &       (5) & (37,74)    & SO                 &  (1) &    (6,28) & (91)       \\
CH$_2$CCH$_2$      &  (1) &        -  & (47)       & NH                 &  (1) &        -  & (39)       & HSO                &  (2) &        -  & (53)       \\
C$_3$H$_5$         &  (2) &        -  & (48)       & NH$_2$             &  (2) &        -  & (39)       & SO$_2$             &  (1) &      (15) & (92)       \\
C$_3$H$_6$         &  (1) &        -  & (49)       & NH$_3$             &  (1) &      (15) & (75)       & HOSO               &  (2) &        -  & (53)       \\
C$_3$H$_7$         &  (1) &        -  & (50)       & N$_2$              &  (1) & (6,18,19) & (76,39)    & SO$_3$             &  (1) &      (15) & (93)       \\
C$_3$H$_8$         &  (1) &        -  & (51)       & N$_2$H             &  (2) &        -  & (55)       & HOSO$_2$           &  (2) &        -  & (53)       \\
C$_4$              &  (1) &        -  & (39)       & N$_2$H$_2$         &  (2) &        -  & (55)       & S$_2$O             &  (1) &        -  & (94)       \\
C$_4$H             &  (2) &        -  & (39)       & H$_2$NN            &  (2) &        -  & (53)       & OCS                &  (1) &      (15) & (39,45)    \\
C$_4$H$_2$         &  (2) &      (15) & (52)       & N$_2$H$_3$         &  (2) &        -  & (55)       & NS                 &  (2) &        -  & (55)       \\
C$_4$H$_3$         &  (2) &        -  & (53)       & N$_2$H$_4$         &  (1) &        -  & (77)       & Si                 &  (1) &       (5) & (37,95)    \\
C$_4$H$_4$         &  (2) &        -  & (54)       & CN                 &  (2) & (6,20,21) & (39)       & SiH                &  (2) &    (6,29) & (39)       \\
C$_4$H$_5$         &  (2) &        -  & (55)       & HCN                &  (1) & (6,22,23) & (78,39)    & SiH$_2$            &  (2) &        -  & (55)       \\
C$_4$H$_6$         &  (2) &        -  & (56)       & HNC                &  (1) &        -  & (39,79)    & SiH$_3$            &  (2) &        -  & (96)       \\
C$_4$H$_7$         &  (2) &        -  & (57)       & H$_2$CN            &  (2) &        -  & (80)       & SiH$_4$            &  (1) &        -  & (97)       \\
C$_4$H$_8$         &  (2) &        -  & (58)       & HCNH               &  (2) &        -  & (53)       & SiC                &  (1) &        -  & (55)       \\
C$_4$H$_9$         &  (2) &        -  & (50)       & CH$_2$NH           &  (3) &        -  & (55)       & SiC$_2$            &  (1) &        -  & (55)       \\
C$_4$H$_{10}$      &  (1) &        -  & (59)       & CH$_2$NH$_2$       &  (2) &        -  & (55)       & SiO                &  (1) &    (6,30) & (39)       \\
C$_5$H$_3$         &  (2) &        -  & (53)       & CH$_3$NH           &  (2) &        -  & (55)       & SiO$_2$            &  (2) &        -  & (53)       \\
C$_5$H$_4$         &  (2) &        -  & (55)       & CH$_3$NH$_2$       &  (2) &        -  & (39)       & SiN                &  (1) &        -  & (53)       \\
C$_5$H$_5$         &  (2) &        -  & (55)       & CH$_2$CN           &  (2) &        -  & (55)       & SiS                &  (1) &    (6,31) & (55)       \\
C$_5$H$_6$         &  (2) &        -  & (55)       & CH$_3$CN           &  (2) &      (15) & (39)       & P                  &  (1) &       (5) & (39,45)    \\
C$_6$H$_2$         &  (1) &        -  & (60)       & NCCN               &  (1) &      (15) & (81)       & PH                 &  (1) &    (6,32) & (39)       \\
C$_6$H$_3$         &  (2) &        -  & (53)       & HC$_3$N            &  (2) &      (15) & (39)       & PH$_2$             &  (1) &        -  & (55)       \\
C$_6$H$_4$         &  (2) &        -  & (55)       & C$_2$H$_2$CN       &  (2) &        -  & (53)       & PH$_3$             &  (1) &      (15) & (98)       \\
C$_6$H$_5$         &  (2) &        -  & (61)       & C$_2$H$_3$CN       &  (2) &        -  & (82)       & P$_2$              &  (1) &        -  & (55)       \\
C$_6$H$_6$         &  (1) &        -  & (62)       & NO                 &  (1) &   (11,24) & (83,39)    & CP                 &  (1) &        -  & (55)       \\
O                  &  (1) &       (5) & (37,63)    & HNO                &  (1) &        -  & (55)       & HCP                &  (2) &        -  & (55)       \\
O($^1$D)           &  (2) &        -  & (45)       & NHOH               &  (2) &        -  & (53)       & PO                 &  (1) &    (6,33) & (55)       \\
OH                 &  (2) &      (11) & (39,45)    & NH$_2$O            &  (2) &        -  & (53)       & PO$_2$             &  (2) &        -  & (55)       \\
H$_2$O             &  (1) &      (11) & (64)       & NH$_2$OH           &  (2) &        -  & (84)       & PN                 &  (1) &    (6,34) & (55)       \\
O$_2$              &  (1) &      (15) & (65)       & NO$_2$             &  (1) &   (11,24) & (85,45)    & Ti                 &  (1) &       (5) & (37,99,45) \\
HO$_2$             &  (2) &      (15) & (39)       & HONO               &  (2) &        -  & (45,68)    & TiO                &  (1) & (6,35,36) & (55)       \\
H$_2$O$_2$         &  (2) &      (15) & (39)       & NO$_3$             &  (1) &        -  & (45,68)    & TiO$_2$            &  (1) &        -  & (55)       \\
O$_3$              &  (1) &      (15) & (39,45)    & HNO$_3$            &  (2) &      (15) & (86,68)    &                    &      &           &            \\
\hline
\end{tabular}
\tablebib{
\tiny
(1) NASA/CEA \citep{McBride2002}, \url{https://www1.grc.nasa.gov/research-and-engineering/ceaweb/};
(2) Third Millenium Thermochemical Database \citep{Goos2018}, \url{https://burcat.technion.ac.il/};
(3) \cite{2018GLA};
(4) \cite{2017SON};
(5) VALD3 \citep{Piskunov1995,Kupka1999,Ryabchikova2015}, \url{https://vald.astro.uu.se/};
(6) ExoMol \citep{Tennyson2024}, \url{https://www.exomol.com/};
(7) \cite{Roueff2019};
(8) \cite{Abgrall1994};
(9) MoLLIST \cite{Bernath2020};
(10) \cite{Masseron2014};
(11) HITEMP \cite{Rothman2010}, \url{https://hitran.org/hitemp/};
(12) \cite{Hargreaves2020};
(13) \cite{Yurchenko2018a};
(14) \cite{McKemmish2020};
(15) HITRAN \cite{Gordon2022}, \url{https://hitran.org/};
(16) \cite{Li2015};
(17) \cite{Somogyi2021};
(18) \cite{Western2018};
(19) \cite{Jans2024};
(20) \cite{Syme2020};
(21) \cite{Syme2021};
(22) \cite{Harris2006};
(23) \cite{Barber2014};
(24) \cite{Hargreaves2019};
(25) \cite{Gorman2019};
(26) \cite{Gomez2024};
(27) \cite{Paulose2015};
(28) \cite{Brady2024};
(29) \cite{Yurchenko2018b};
(30) \cite{Yurchenko2022};
(31) \cite{Upadhyay2018};
(32) \cite{Langleben2019};
(33) \cite{Prajapat2017};
(34) \cite{Semenov2025};
(35) \cite{McKemmish2019};
(36) \cite{McKemmish2024};
(37) NORAD-Atomic-Data \citep{Nahar2020,Nahar2024}, \url{https://norad.astronomy.osu.edu/};
(38) \cite{Nahar2021};
(39) Leiden database \citep{Heays2017,Hrodmarsson2023}, \url{https://home.strw.leidenuniv.nl/~ewine/photo/};
(40) \cite{Nahar2010};
(41) \cite{Nahar1991};
(42) \cite{Khamaganov2007}, \cite{Macpherson1985}, \cite{Arthur1986}, \cite{Herzberg1961}, \cite{Kassner1994}, \cite{Wilson1994}, \cite{North1995}, \cite{Wu2004}, \cite{Gans2010}, \cite{Chupka1968}, \cite{Litorja1997}, \cite{Loison2010}; 
(43) \cite{Au1993}, \cite{Kameta2002}, \cite{Gans2011}; 
(44) \cite{Cooper1995a}, \cite{Hudson1971}; 
(45) PHID database \citep{Huebner1992,Huebner2015}, \url{https://phidrates.space.swri.edu/};
(46) \cite{Ho1998}, \cite{Chen2000}. 
}
\end{table}

\noindent
\begin{table}
\centering
\begin{tabular}{c}
\multicolumn{1}{c}{~~~~~~~~~~~~~~~~~~~~~~~~~~~~~~~~~~~~~~~~~~~~~~~~~~~~~~~~~~~~~~~~~~~~~~~~~~~~~~~~~~~~~~~~~~~~~~~~~~~~~~~~~~~~~~~~~~~~~~~~~~~~~~~~~~~~~~~~~~~~~~~~~~~~~~~~~~~~~~~~~~~~~~~~~~~~~~~~~~~~~~~~~~~~~~~~~~~~~~~~~~~~~} \\
\hline
\end{tabular}
\tablebib{
\tiny
(47) \cite{Holland1999}, \cite{Chen2000}; 
(48) \cite{Jenkin1993}; 
(49) \cite{Koizumi1985}, \cite{Christianson2021}; 
(50) \cite{Wendt1984}; 
(51) \cite{Au1993}; 
(52) \cite{Ferradaz2009}, \cite{Fahr1994}; 
(53) guess cross section of 1 Mb adopted in the wavelength range 100-250 nm;
(54) \cite{Fahr1996}; 
(55) guess cross section of 1 Mb adopted from photoionization threshold to 250 nm;
(56) \cite{Schoen1962}, \cite{Palmer2010}; 
(57) \cite{Bayratceken2006}; 
(58) \cite{Koizumi1985}, \cite{Gary1954}; 
(59) \cite{Raymonda1967}, \cite{Doner2023}; 
(60) \cite{Kloster-Jensen1974}, \cite{Shindo2003}; 
(61) \cite{Wallington1998b}; 
(62) \cite{Feng2002}, \cite{Boechat-Roberty2004}; 
(63) \cite{Nahar1998};
(64) \cite{Chan1993a}, \cite{Fillion2003}, \cite{Fillion2004}, \cite{Mota2005}, \cite{Slanger1982}, \cite{Crovisier1989}; 
(65) \cite{Chan1993b}, \cite{Holland1993}, \cite{Yoshino2005}, \cite{Lee1977b}; 
(66) \cite{Bruna1976}, \cite{2012VEN}, \cite{Hochanadel1980}, \cite{Loison1991}; 
(67) \cite{Cooper1996}; 
(68) \cite{Huestis2010}; 
(69) \cite{Sander2011}; 
(70) \cite{Rabalais1971}, \cite{Laufer1971}; 
(71) \cite{Cameron2002}, \cite{Maricq1996}, \cite{Rajakumar2007}; 
(72) \cite{Anastasi1990}; 
(73) data from Maricq (1999) at The MPI-Mainz UV/VIS Spectral Atlas of Gaseous Molecules of Atmospheric Interest \citep{Keller-Rudek2013}, \url{https://www.uv-vis-spectral-atlas-mainz.org/uvvis/}; 
(74) \cite{Nahar1997};
(75) \cite{Burton1993}, \cite{Edvardsson1999}, \cite{Wu2007b}, \cite{Chen2006}, \cite{McNesby1962}, \cite{Okabe1967}, \cite{Groth1968}, \cite{Lilly1973}, \cite{Slanger1982}; 
(76) \cite{Chan1993c}; 
(77) \cite{Biehl1991}, \cite{Vaghjiani1993}; 
(78) \cite{Nuth1982}; 
(79) \cite{Gans2019}; 
(80) \cite{Horne1970}; 
(81) \cite{Nuth1982}, \cite{Halpern2018}, LISA Titan Spectroscopic Database, \url{http://www.lisa.u-pec.fr/GPCOS/SCOOPweb/}; 
(82) \cite{Eden2003}; 
(83) \cite{Watanabe1967}, \cite{Guest1981}, \cite{Chan1993d}; 
(84) \cite{Betts1965}; 
(85) \cite{Au1997}; 
(86) \cite{Okabe1980}; 
(87) \cite{Chan1994}; 
(88) \cite{Stockwell1978}; 
(89) \cite{Verner1996};
(90) \cite{Feng1999a}; 
(91) \cite{Nee1986}, \cite{Phillips1981}, \cite{Norwood1989}; 
(92) \cite{Feng1999b}, \cite{Manatt1993}, \cite{Holland1995}, \cite{Wu2000}, \cite{Driscoll1968}; 
(93) \cite{Burkholder2015}; 
(94) \cite{Frandsen2020}; 
(95) \cite{Nahar2000};
(96) \cite{Baklanov2001}; 
(97) \cite{Cooper1995b}; 
(98) \cite{Zarate1990}, \cite{Carravetta1999}; 
(99) \cite{Nahar2015}.
}
\end{table}

\begin{table}
\tiny
\caption{Reaction list (only a sample is shown).}
\label{tab:reactions}
\centering
\begin{tabular}{
l
@{\hspace{0.2cm}}c@{\hspace{0.1cm}}r@{\hspace{0.1cm}}r
@{\hspace{0.1cm}}c
@{\hspace{0.1cm}}c@{\hspace{0.1cm}}r@{\hspace{0.1cm}}r
@{\hspace{0.1cm}}c
@{\hspace{0.1cm}}c@{\hspace{0.1cm}}c
@{\hspace{0.1cm}}c
@{\hspace{0.1cm}}c 
@{\hspace{0.1cm}}c
@{\hspace{0.1cm}}c
@{\hspace{0.1cm}}c
@{\hspace{0.1cm}}c
@{\hspace{0.1cm}}c
@{\hspace{0.1cm}}c
@{\hspace{0.1cm}}c
}
\hline \hline
\\
\multicolumn{20}{c}{Bimolecular reactions} \\
\\
\hline
Reaction &
\multicolumn{1}{c}{$\alpha$} & \multicolumn{1}{c}{$\beta$} & \multicolumn{1}{c}{$\gamma$} &
&
& & &
&
& &
&
&
&
\multicolumn{3}{c}{Temperature range} &
&
Error & Ref. \\
&
\multicolumn{1}{c}{(cm$^3$ s$^{-1}$)} & & \multicolumn{1}{c}{(K)} &
&
& & &
&
& &
&
&
&
\multicolumn{3}{c}{(K)} &
&
& \\
\hline
 H + CH $\rightarrow$ C + H$_2$                       & 2.0\,$\times$\,10$^{-10}$  &    0.0   &       0 & &                            &         &        & &       &                             & &       & &            &  1500-2500 &          & & A & (62) \\
 O$_2$ + C$_4$H$_5$ $\rightarrow$ HO$_2$ + C$_4$H$_4$ & 1.2\,$\times$\,10$^{-12}$  &    0.21  &    6420 & &                            &         &        & &       &                             & &       & &            &   300-2000 &          & & B & (748) \\
 HCCO + CH $\rightarrow$ CO + C$_2$H$_2$              & 8.3\,$\times$\,10$^{-11}$  &    0.0   &       0 & &                            &         &        & &       &                             & &       & &            &   300-2000 &          & & C & (172) \\
 NH$_2$ + OH $\rightarrow$ NH + H$_2$O                & 3.6\,$\times$\,10$^{-13}$  &    1.97  & $-$1130 & &                            &         &        & &       &                             & &       & &            &   300-3000 &          & & A & (463) \\
 H$_2$S + CH $\rightarrow$ H$_2$CS + H                & 2.8\,$\times$\,10$^{-10}$  &    0.0   &       0 & &                            &         &        & &       &                             & &       & &            &   330-330  &          & & A & (267) \\
\hline
\\
\multicolumn{20}{c}{Pressure-dependent reactions} \\
\\
\hline
&
\multicolumn{3}{c}{$k_0$} &
& 
\multicolumn{3}{c}{$k_\infty$} &
&
\multicolumn{2}{c}{$F_c$} &
&
&
& 
\multicolumn{1}{c}{$k_0$} & \multicolumn{1}{c}{$k_\infty$} & \multicolumn{1}{c}{$F_c$} &
&
& \\
\cline{2-4} \cline{6-8} \cline{10-11} \cline{15-17}
Reaction &
\multicolumn{1}{c}{$\alpha$} & \multicolumn{1}{c}{$\beta$} & \multicolumn{1}{c}{$\gamma$} &
&
\multicolumn{1}{c}{$\alpha$} & \multicolumn{1}{c}{$\beta$} & \multicolumn{1}{c}{$\gamma$} &
&
A & B &
&
M &
&
\multicolumn{3}{c}{Temperature range (K)} &
&
Error & Ref. \\
\hline
 H + CH$_3$ + M = CH$_4$ + M                          & 5.9\,$\times$\,10$^{-29}$  & $-$1.8   &       0 & &  3.5\,$\times$\,10$^{-10}$ &    0.0  &      0 & & 0.611 & $-$1.4\,$\times$\,10$^{-4}$ & &    Ar & &   300-1000 &   300-2000 & 300-2000 & & A & (62) \\
 CH$_3$ + CH$_3$ + M $\rightarrow$ C$_2$H$_6$ + M     & 1.6\,$\times$\,10$^{-24}$  & $-$7.0   &    1390 & &  6.0\,$\times$\,10$^{-11}$ &    0.0  &      0 & & 0.516 & $-$2.2\,$\times$\,10$^{-4}$ & &    Ar & &   300-2000 &   300-2000 & 300-2000 & & A & (62) \\
 O$_2$ + H + M $\rightarrow$ HO$_2$ + M               & 4.8\,$\times$\,10$^{-32}$  & $-$1.29  &       0 & &  5.3\,$\times$\,10$^{-11}$ &  0.604  & $-$121 & & 0.458 &                             & & N$_2$ & &   300-2000 &   300-2000 &          & & A & (976) \\
 N + O + M $\rightarrow$ NO + M                       & 5.5\,$\times$\,10$^{-33}$  &    0.0   &  $-$155 & &                            &         &        & &       &                             & & N$_2$ & &    196-298 &            &          & & C & (131) \\
 S + H$_2$S + M $\rightarrow$ H$_2$S$_2$ + M          & 6.7\,$\times$\,10$^{-31}$  & $-$1.612 &     840 & &  1.6\,$\times$\,10$^{-13}$ &    1.28 & $-$240 & & 0.810 & $-$5.8\,$\times$\,10$^{-4}$ & &    Ar & &   291-1042 &   291-1042 & 291-1042 & & A & (290) \\
\hline
\end{tabular}
\tablebib{
\tiny
(1)\,\cite{2015ABE}; 
(2)\,\cite{2015ABI}; 
(3)\,\cite{1981ADA}; 
(4)\,\cite{1973ADE}; 
(5)\,\cite{1993ADU}; 
(6)\,\cite{1994ADU}; 
(7)\,\cite{1996ADU}; 
(8)\,\cite{2007AGU}; 
(9)\,\cite{2023AGU}; 
(10)\,\cite{2019ALA}; 
(11)\,\cite{2022ALA}; 
(12)\,\cite{2002ALB}; 
(13)\,\cite{1989ALEa}; 
(14)\,\cite{1989ALEb}; 
(15)\,\cite{1994ALE}; 
(16)\,\cite{2015ALI}; 
(17)\,\cite{2016ALI}; 
(18)\,\cite{2015ALT}; 
(19)\,\cite{2000ALZ}; 
(20)\,\cite{2021ALZ}; 
(21)\,\cite{1984ANA}; 
(22)\,\cite{1988ANA}; 
(23)\,\cite{2018AND}; 
(24)\,\cite{2004ANG}; 
(25)\,\cite{2018ANG}; 
(26)\,\cite{1981ARA}; 
(27)\,\cite{2020ARA}; 
(28)\,\cite{1983ARI}; 
(29)\,\cite{1971ARM}; 
(30)\,\cite{1978ART}; 
(31)\,\cite{1997ARTa}; 
(32)\,\cite{1997ARTb}; 
(33)\,\cite{1971ASA}; 
(34)\,\cite{2018ASG}; 
(35)\,\cite{1962ASH}; 
(36)\,\cite{2018ASS}; 
(37)\,\cite{1992ATA}; 
(38)\,\cite{1973ATK}; 
(39)\,\cite{1989ATK}; 
(40)\,\cite{1997ATK}; 
(41)\,\cite{2004ATK}; 
(42)\,\cite{2006ATK}; 
(43)\,\cite{2001BAC}; 
(44)\,\cite{2014BAD}; 
(45)\,\cite{1987BAG}; 
(46)\,\cite{1990BAG}; 
(47)\,\cite{2021BAI}; 
(48)\,\cite{1965BAK}; 
(49)\,\cite{2008BAL}; 
(50)\,\cite{2009BAL}; 
(51)\,\cite{2015BAL}; 
(52)\,\cite{2023BAP}; 
(53)\,\cite{1991BAR}; 
(54)\,\cite{2020BAR}; 
(55)\,\cite{1988BASa}; 
(56)\,\cite{1988BASb}; 
(57)\,\cite{2011BAT}; 
(58)\,\cite{1971BAU}; 
(59)\,\cite{1985BAU}; 
(60)\,\cite{1992BAU}; 
(61)\,\cite{1994BAU}; 
(62)\,\cite{2005BAU}; 
(63)\,\cite{1992BECa}; 
(64)\,\cite{1992BECb}; 
(65)\,\cite{1993BEC}; 
(66)\,\cite{1995BEC}; 
(67)\,\cite{1997BEC}; 
(68)\,\cite{1999BEC}; 
(69)\,\cite{2000BEC}; 
(70)\,\cite{2005BECa}; 
(71)\,\cite{2005BECb}; 
(72)\,\cite{2017BED}; 
(73)\,\cite{2022BED}; 
(74)\,\cite{1994BEN}; 
(75)\,\cite{2007BEN}; 
(76)\,\cite{1973BER}; 
(77)\,\cite{1999BER}; 
(78)\,\cite{2001BER}; 
(79)\,\cite{2008BER}; 
(80)\,\cite{2009BER}; 
(81)\,\cite{2010BERa}; 
(82)\,\cite{2010BERb}; 
(83)\,\cite{2019BIA}; 
(84)\,\cite{1994BIG}; 
(85)\,\cite{1995BIG}; 
(86)\,\cite{1957BIR}; 
(87)\,\cite{2020BLA}; 
(88)\,\cite{2000BLI}; 
(89)\,\cite{2003BLI}; 
(90)\,\cite{2004BLI}; 
(91)\,\cite{2006BLI}; 
(92)\,\cite{2010BLI}; 
(93)\,\cite{2011BLI}; 
(94)\,\cite{2012BLI}; 
(95)\,\cite{1996BOC}; 
(96)\,\cite{2004BOG}; 
(97)\,\cite{1985BOH}; 
(98)\,\cite{1988BOH}; 
(99)\,\cite{1987BOO}; 
(100)\,\cite{1983BOS}; 
(101)\,\cite{1997BOU}; 
(102)\,\cite{2012BOU}; 
(103)\,\cite{2013BOU}; 
(104)\,\cite{2015BOU}; 
(105)\,\cite{2019BOUa}; 
(106)\,\cite{2019BOUb}; 
(107)\,\cite{2023BOU}; 
(108)\,\cite{2018BOW}; 
(109)\,\cite{2020BOW}; 
(110)\,\cite{1990BOZ}; 
(111)\,\cite{1995BOZ}; 
(112)\,\cite{1967BRA}; 
(113)\,\cite{2002BRA}; 
(114)\,\cite{1973BRO}; 
(115)\,\cite{1975BRU}; 
(116)\,\cite{1983BRU}; 
(117)\,\cite{2009BRU}; 
(118)\,\cite{1993BRY}; 
(119)\,\cite{1990BUC}; 
(120)\,\cite{1994BUC}; 
(121)\,\cite{1995BUD}; 
(122)\,\cite{2020BUE}; 
(123)\,\cite{2021BUL}; 
(124)\,\cite{2004BUN}; 
(125)\,\cite{1997BURa}; 
(126)\,\cite{1997BURb}; 
(127)\,\cite{2018BUR}; 
(128)\,\cite{2020BUR}; 
(129)\,\cite{1981BUT}; 
(130)\,\cite{2005BUT}; 
(131)\,\cite{1973CAM}; 
(132)\,\cite{1993CAM}; 
(133)\,\cite{2001CAM}; 
(134)\,\cite{2022CAM}; 
(135)\,\cite{1988CAN}; 
(136)\,\cite{1991CAN}; 
(137)\,\cite{1997CAN}; 
(138)\,\cite{2001CAN}; 
(139)\,\cite{2001CARa}; 
(140)\,\cite{2001CARb}; 
(141)\,\cite{2002CAR}; 
(142)\,\cite{2003CARa}; 
(143)\,\cite{2003CARb}; 
(144)\,\cite{2003CARc}; 
(145)\,\cite{2005CARa}; 
(146)\,\cite{2005CARb}; 
(147)\,\cite{2005CARc}; 
(148)\,\cite{2006CAR}; 
(149)\,\cite{2007CAR}; 
(150)\,\cite{2012CAR}; 
(151)\,\cite{2013CAR}; 
(152)\,\cite{2019CAS}; 
(153)\,\cite{2021CAS}; 
(154)\,\cite{2021CER}; 
(155)\,\cite{2022CER}; 
(156)\,\cite{1986CHA}; 
(157)\,\cite{1995CHAb}; 
(158)\,\cite{1998CHA}; 
(159)\,\cite{1999CHAa}; 
(160)\,\cite{1999CHAb}; 
(161)\,\cite{2000CHA}; 
(162)\,\cite{2001CHA}; 
(163)\,\cite{2007CHA}; 
(164)\,\cite{2017CHA}; 
(165)\,\cite{2012CHE}; 
(166)\,\cite{2016CHE}; 
(167)\,\cite{2019CHE}; 
(168)\,\cite{2000CHO}; 
(169)\,\cite{2004CHOa}; 
(170)\,\cite{2004CHOb}; 
(171)\,\cite{2005CHO}; 
(172)\,\cite{2017CHR}; 
(173)\,\cite{1975CHU}; 
(174)\,\cite{2020CHU}; 
(175)\,\cite{2006CIM}; 
(176)\,\cite{1967CLY}; 
(177)\,\cite{1982CLY}; 
(178)\,\cite{1985COB}; 
(179)\,\cite{2023COB}; 
(180)\,\cite{1991COH}; 
(181)\,\cite{1989COL}; 
(182)\,\cite{2019COL}; 
(183)\,\cite{2021COL}; 
(184)\,\cite{2003COO}; 
(185)\,\cite{2020COX}; 
(186)\,\cite{2001CRE}; 
(187)\,\cite{2006CUR}; 
(188)\,\cite{2009DAS}; 
(189)\,\cite{2013DAS}; 
(190)\,\cite{1995DAE}; 
(191)\,\cite{2020DAI}; 
(192)\,\cite{2007DAM}; 
(193)\,\cite{2014DAM}; 
(194)\,\cite{2001DAN}; 
(195)\,\cite{2019DANa}; 
(196)\,\cite{2019DANb}; 
(197)\,\cite{1995DAR}; 
(198)\,\cite{2012DARa}; 
(199)\,\cite{2012DARb}; 
(200)\,\cite{2013DAR}; 
(201)\,\cite{2005DAU}; 
(202)\,\cite{2008DAU}; 
(203)\,\cite{1990DAV}; 
(204)\,\cite{1991DAV}; 
(205)\,\cite{1999DAV}; 
(206)\,\cite{1992DEA}; 
(207)\,\cite{2000DEA}; 
(208)\,\cite{2001DEC}; 
(209)\,\cite{2003DEC}; 
(210)\,\cite{2006DEE}; 
(211)\,\cite{1980DEM}; 
(212)\,\cite{1982DEM}; 
(213)\,\cite{1997DEM}; 
(214)\,\cite{2008DEN}; 
(215)\,\cite{1998DEP}; 
(216)\,\cite{2003DES}; 
(217)\,\cite{1994DIA}; 
(218)\,\cite{1995DIAa}; 
(219)\,\cite{1995DIAc}; 
(220)\,\cite{1993DOB}; 
(221)\,\cite{2016DOB}; 
(222)\,\cite{2021DOD}; 
(223)\,\cite{2007DOL}; 
(224)\,\cite{2005DON}; 
(225)\,\cite{2018DOU}; 
(226)\,\cite{2019DOUa}; 
(227)\,\cite{2019DOUb}; 
(228)\,\cite{2020DOU}; 
(229)\,\cite{2022DOU}; 
(230)\,\cite{1985DRA}; 
(231)\,\cite{2004DU}; 
(232)\,\cite{2005DU}; 
(233)\,\cite{2008DU}; 
(234)\,\cite{2012DU}; 
(235)\,\cite{2013DUA}; 
(236)\,\cite{1971DUR}; 
(237)\,\cite{1988DUR}; 
(238)\,\cite{1989DUR}; 
(239)\,\cite{2011DUR}; 
(240)\,\cite{2013DUT}; 
(241)\,\cite{1992EDE}; 
(242)\,\cite{1966EDW}; 
(243)\,\cite{1998EDW}; 
(244)\,\cite{1987EHB}; 
(245)\,\cite{1975ENG}; 
(246)\,\cite{2019ESK}; 
(247)\,\cite{2016ESS}; 
(248)\,\cite{1993FAG}; 
(249)\,\cite{1990FAH}; 
(250)\,\cite{1991FAH}; 
(251)\,\cite{1969FAI}; 
(252)\,\cite{2000FAR}; 
(253)\,\cite{2013FAR}; 
(254)\,\cite{2006FENb}; 
(255)\,\cite{2007FEN}; 
(256)\,\cite{2008FEN}; 
(257)\,\cite{2011FEN}; 
(258)\,\cite{2013FEN}; 
(259)\,\cite{1998FERb}; 
(260)\,\cite{2001FER}; 
(261)\,\cite{2006FER}; 
(262)\,\cite{1975FIF}; 
(263)\,\cite{1976FIF}; 
(264)\,\cite{1964FIS}; 
(265)\,\cite{1998FIT}; 
(266)\,\cite{2019FIT}; 
(267)\,\cite{2002FLE}; 
(268)\,\cite{2001FON}; 
(269)\,\cite{2006FON}; 
(270)\,\cite{2024FOR}; 
(271)\,\cite{1999FOU}; 
(272)\,\cite{1980FRA}; 
(273)\,\cite{1986FRA}; 
(274)\,\cite{1982FRI}; 
(275)\,\cite{2004FRI}; 
(276)\,\cite{1993FRO}; 
(277)\,\cite{1997FULa}; 
(278)\,\cite{1997FULb}; 
(279)\,\cite{2019FUL}; 
(280)\,\cite{2001GAL}; 
(281)\,\cite{2003GAL}; 
(282)\,\cite{2018GAL}; 
(283)\,\cite{2023GAL}; 
(284)\,\cite{2007GAN}; 
(285)\,\cite{2008GAN}; 
(286)\,\cite{2010GAN}; 
(287)\,\cite{2006GAO}; 
(288)\,\cite{2010GAO}; 
(289)\,\cite{2011GAOa}; 
(290)\,\cite{2011GAOb}; 
(291)\,\cite{2014GAO}; 
(292)\,\cite{2018GAO}; 
(293)\,\cite{2020GAO}; 
(294)\,\cite{1990GAR}; 
(295)\,\cite{1998GAR}; 
(296)\,\cite{2012GAR}; 
(297)\,\cite{2021GARa}; 
(298)\,\cite{2021GARb}; 
(299)\,\cite{1975GAU}; 
(300)\,\cite{1987GEE}; 
(301)\,\cite{1971GEH}; 
(302)\,\cite{2000GEP}; 
(303)\,\cite{2004GEP}; 
(304)\,\cite{1967GET}; 
(305)\,\cite{1988GHI}; 
(306)\,\cite{2017GHI}.
}
\end{table}

\noindent
\begin{table}
\centering
\begin{tabular}{c}
\multicolumn{1}{c}{~~~~~~~~~~~~~~~~~~~~~~~~~~~~~~~~~~~~~~~~~~~~~~~~~~~~~~~~~~~~~~~~~~~~~~~~~~~~~~~~~~~~~~~~~~~~~~~~~~~~~~~~~~~~~~~~~~~~~~~~~~~~~~~~~~~~~~~~~~~~~~~~~~~~~~~~~~~~~~~~~~~~~~~~~~~~~~~~~~~~~~~~~~~~~~~~~~~~~~~~~~~~~} \\
\hline
\end{tabular}
\tablebib{
\tiny
(307)\,\cite{2017GHO}; 
(308)\,\cite{2010GIE}; 
(309)\,\cite{2016GIMa}; 
(310)\,\cite{2016GIMb}; 
(311)\,\cite{2022GIR}; 
(312)\,\cite{1975GLA}; 
(313)\,\cite{1986GLA}; 
(314)\,\cite{1996GLAa}; 
(315)\,\cite{1996GLAb}; 
(316)\,\cite{2013GLA}; 
(317)\,\cite{2014GLA}; 
(318)\,\cite{2015GLA}; 
(319)\,\cite{2018GLA}; 
(320)\,\cite{2020GLA}; 
(321)\,\cite{1983GOL}; 
(322)\,\cite{2009GOL}; 
(323)\,\cite{2011GOL}; 
(324)\,\cite{2009GOM}; 
(325)\,\cite{2023GOMa}; 
(326)\,\cite{2023GOMb}; 
(327)\,\cite{2024GOMa}; 
(328)\,\cite{2024GOMb}; 
(329)\,\cite{2012GON}; 
(330)\,\cite{2022GON}; 
(331)\,\cite{2024GON}; 
(332)\,\cite{2012GOO}; 
(333)\,\cite{1999GOU}; 
(334)\,\cite{2003GOU}; 
(335)\,\cite{2007GOU}; 
(336)\,\cite{2009GOU}; 
(337)\,\cite{2013GOU}; 
(338)\,\cite{1965GRA}; 
(339)\,\cite{1966GRA}; 
(340)\,\cite{1978GRA}; 
(341)\,\cite{1988GRO}; 
(342)\,\cite{1989GRO}; 
(343)\,\cite{2007GU}; 
(344)\,\cite{1995GUO}; 
(345)\,\cite{2007GUO}; 
(346)\,\cite{2017GUO}; 
(347)\,\cite{2019GUP}; 
(348)\,\cite{1986HAC}; 
(349)\,\cite{1994HAC}; 
(350)\,\cite{2005HAC}; 
(351)\,\cite{2001HAH}; 
(352)\,\cite{1993HAIa}; 
(353)\,\cite{1993HAIb}; 
(354)\,\cite{1973HAL}; 
(355)\,\cite{1984HAN}; 
(356)\,\cite{1992HAN}; 
(357)\,\cite{2011HAN}; 
(358)\,\cite{1993HAR}; 
(359)\,\cite{2005HARa}; 
(360)\,\cite{2005HARb}; 
(361)\,\cite{2007HAR}; 
(362)\,\cite{1990HAS}; 
(363)\,\cite{2019HAS}; 
(364)\,\cite{2020HAS}; 
(365)\,\cite{2003HAW}; 
(366)\,\cite{1988HE}; 
(367)\,\cite{1992HEa}; 
(368)\,\cite{1992HEb}; 
(369)\,\cite{1993HE}; 
(370)\,\cite{2020HE}; 
(371)\,\cite{2013HEB}; 
(372)\,\cite{1995HEN}; 
(373)\,\cite{1988HER}; 
(374)\,\cite{2013HIC}; 
(375)\,\cite{2014HIC}; 
(376)\,\cite{2015HIC}; 
(377)\,\cite{2016HIC}; 
(378)\,\cite{2022HICa}; 
(379)\,\cite{2022HICb}; 
(380)\,\cite{2024HIC}; 
(381)\,\cite{1990HID}; 
(382)\,\cite{2008HIG}; 
(383)\,\cite{1987HIL}; 
(384)\,\cite{2007HIN}; 
(385)\,\cite{1990HIP}; 
(386)\,\cite{1986HJO}; 
(387)\,\cite{2021HOB}; 
(388)\,\cite{1970HOH}; 
(389)\,\cite{1983HOM}; 
(390)\,\cite{2010HON}; 
(391)\,\cite{2012HON}; 
(392)\,\cite{1997HOOa}; 
(393)\,\cite{1997HOOb}; 
(394)\,\cite{2000HOU}; 
(395)\,\cite{2005HOU}; 
(396)\,\cite{1997HOV}; 
(397)\,\cite{2009HOY}; 
(398)\,\cite{2011HOY}; 
(399)\,\cite{1982HSU}; 
(400)\,\cite{2013HU}; 
(401)\,\cite{2017HUAa}; 
(402)\,\cite{2017HUAb}; 
(403)\,\cite{2018HUA}; 
(404)\,\cite{2011HUO}; 
(405)\,\cite{2018HUO}; 
(406)\,\cite{1971HUS}; 
(407)\,\cite{1975HUS}; 
(408)\,\cite{1977HUS}; 
(409)\,\cite{1978HUSa}; 
(410)\,\cite{1978HUSb}; 
(411)\,\cite{1999HUS}; 
(412)\,\cite{2003HWA}; 
(413)\,\cite{2010HWA}; 
(414)\,\cite{1996IKE}; 
(415)\,\cite{1961IMA}; 
(416)\,\cite{1999INO}; 
(417)\,\cite{2007ISM}; 
(418)\,\cite{2009ISM}; 
(419)\,\cite{2017IZA}; 
(420)\,\cite{1965JAC}; 
(421)\,\cite{1974JAF}; 
(422)\,\cite{2015JAN}; 
(423)\,\cite{1988JAS}; 
(424)\,\cite{2007JASa}; 
(425)\,\cite{2007JASb}; 
(426)\,\cite{2009JAS}; 
(427)\,\cite{2003JAV}; 
(428)\,\cite{2005JAY}; 
(429)\,\cite{2004JIM}; 
(430)\,\cite{2020JIN}; 
(431)\,\cite{1995JOD}; 
(432)\,\cite{1966JOH}; 
(433)\,\cite{1986JOH}; 
(434)\,\cite{2000JOH}; 
(435)\,\cite{2006JOS}; 
(436)\,\cite{2007JU}; 
(437)\,\cite{2002KAI}; 
(438)\,\cite{1979KAJ}; 
(439)\,\cite{2011KAP}; 
(440)\,\cite{1999KAR}; 
(441)\,\cite{2004KAR}; 
(442)\,\cite{1983KAY}; 
(443)\,\cite{1959KEN}; 
(444)\,\cite{1972KER}; 
(445)\,\cite{1988KER}; 
(446)\,\cite{2007KER}; 
(447)\,\cite{2015KER}; 
(448)\,\cite{2017KHA}; 
(449)\,\cite{2019KHA}; 
(450)\,\cite{1988KIE}; 
(451)\,\cite{1992KIE}; 
(452)\,\cite{1993KIE}; 
(453)\,\cite{1994KIE}; 
(454)\,\cite{1996KIE}; 
(455)\,\cite{2003KIM}; 
(456)\,\cite{1994KIN}; 
(457)\,\cite{2002KIS}; 
(458)\,\cite{1995KLA}; 
(459)\,\cite{1974KLE}; 
(460)\,\cite{2005KLI}; 
(461)\,\cite{2006KLI}; 
(462)\,\cite{2009KLIa}; 
(463)\,\cite{2009KLIb}; 
(464)\,\cite{2010KLI}; 
(465)\,\cite{2011KLIa}; 
(466)\,\cite{2011KLIb}; 
(467)\,\cite{2013KLI}; 
(468)\,\cite{2015KLI}; 
(469)\,\cite{2017KLI}; 
(470)\,\cite{2022KLI}; 
(471)\,\cite{1996KNY}; 
(472)\,\cite{2001KNY}; 
(473)\,\cite{2017KNY}; 
(474)\,\cite{1986KOD}; 
(475)\,\cite{2001KONa}; 
(476)\,\cite{2001KONb}; 
(477)\,\cite{2020KOV}; 
(478)\,\cite{1994KRA}; 
(479)\,\cite{2004KRA}; 
(480)\,\cite{2007KRA}; 
(481)\,\cite{2017KRA}; 
(482)\,\cite{1997KRU}; 
(483)\,\cite{1999KRU}; 
(484)\,\cite{1995KUK}; 
(485)\,\cite{2001KUN}; 
(486)\,\cite{1995KUR}; 
(487)\,\cite{2010KUR}; 
(488)\,\cite{1961LAI}; 
(489)\,\cite{1962LAI}; 
(490)\,\cite{2019LAK}; 
(491)\,\cite{2021LAK}; 
(492)\,\cite{1968LAM}; 
(493)\,\cite{1994LAN}; 
(494)\,\cite{1995LAN}; 
(495)\,\cite{2008LAN}; 
(496)\,\cite{1983LAU}; 
(497)\,\cite{2004LAU}; 
(498)\,\cite{1998LEP}; 
(499)\,\cite{2002LEP}; 
(500)\,\cite{1977LEE}; 
(501)\,\cite{2005LEE}; 
(502)\,\cite{2012LEEa}; 
(503)\,\cite{2012LEEb}; 
(504)\,\cite{2024LEE}; 
(505)\,\cite{1989LEI}; 
(506)\,\cite{2012LEO}; 
(507)\,\cite{1986LES}; 
(508)\,\cite{2002LI}; 
(509)\,\cite{2004LI}; 
(510)\,\cite{2006LI}; 
(511)\,\cite{2014LI}; 
(512)\,\cite{2017LIa}; 
(513)\,\cite{2017LIb}; 
(514)\,\cite{2020LIa}; 
(515)\,\cite{2020LIb}; 
(516)\,\cite{1993LIF}; 
(517)\,\cite{1997LIF}; 
(518)\,\cite{1998LIF}; 
(519)\,\cite{1992LIN}; 
(520)\,\cite{1996LIN}; 
(521)\,\cite{1973LIS}; 
(522)\,\cite{1995LIS}; 
(523)\,\cite{1988LIU}; 
(524)\,\cite{2002LIUa}; 
(525)\,\cite{2002LIUb}; 
(526)\,\cite{2006LIU}; 
(527)\,\cite{2019LIU}; 
(528)\,\cite{2013LOC}; 
(529)\,\cite{2004LOI}; 
(530)\,\cite{2006LOI}; 
(531)\,\cite{2012LOI}; 
(532)\,\cite{2014LOIa}; 
(533)\,\cite{2014LOIb}; 
(534)\,\cite{2015LOIa}; 
(535)\,\cite{2015LOIb}; 
(536)\,\cite{2017LOI}; 
(537)\,\cite{1984LOU}; 
(538)\,\cite{2000LU}; 
(539)\,\cite{2003LU}; 
(540)\,\cite{2004LU}; 
(541)\,\cite{2005LU}; 
(542)\,\cite{2006LU}; 
(543)\,\cite{2023LUC}; 
(544)\,\cite{2006LUD}; 
(545)\,\cite{2015LYN}; 
(546)\,\cite{2005MAC}; 
(547)\,\cite{2007MAC}; 
(548)\,\cite{2019MAC}; 
(549)\,\cite{1997MAD}; 
(550)\,\cite{2020MAF}; 
(551)\,\cite{2019MAI}; 
(552)\,\cite{2020MAI}; 
(553)\,\cite{2022MAI}; 
(554)\,\cite{2011MAK}; 
(555)\,\cite{2021MAL}; 
(556)\,\cite{1999MAN}; 
(557)\,\cite{2018MAN}; 
(558)\,\cite{2020MAO}; 
(559)\,\cite{1931MAR}; 
(560)\,\cite{1968MAR}; 
(561)\,\cite{1983MAR}; 
(562)\,\cite{1989MARa}; 
(563)\,\cite{1989MARb}; 
(564)\,\cite{1992MAR}; 
(565)\,\cite{1999MAR}; 
(566)\,\cite{2012MAR}; 
(567)\,\cite{2019MAR}; 
(568)\,\cite{2020MAS}; 
(569)\,\cite{1996MAT}; 
(570)\,\cite{2010MAT}; 
(571)\,\cite{2011MAT}; 
(572)\,\cite{2014MAT}; 
(573)\,\cite{1966MAY}; 
(574)\,\cite{1967MAY}; 
(575)\,\cite{2018MAZ}; 
(576)\,\cite{2011MEA}; 
(577)\,\cite{1996MEBa}; 
(578)\,\cite{1996MEBb}; 
(579)\,\cite{1996MEBc}; 
(580)\,\cite{1997MEB}; 
(581)\,\cite{1998MEBa}; 
(582)\,\cite{1998MEBb}; 
(583)\,\cite{2006MEB}; 
(584)\,\cite{2017MEB}; 
(585)\,\cite{2020MED}; 
(586)\,\cite{1996MEH}; 
(587)\,\cite{1992MEL}; 
(588)\,\cite{2018MEN}; 
(589)\,\cite{2019MEN}; 
(590)\,\cite{2024MEN}; 
(591)\,\cite{1989MER}; 
(592)\,\cite{1991MER}; 
(593)\,\cite{2018MER}; 
(594)\,\cite{1983MET}; 
(595)\,\cite{2000MEY}; 
(596)\,\cite{2005MEY}; 
(597)\,\cite{1974MIC}; 
(598)\,\cite{1993MICa}; 
(599)\,\cite{1993MICb}; 
(600)\,\cite{1994MICa}; 
(601)\,\cite{1994MICb}; 
(602)\,\cite{2005MIC}; 
(603)\,\cite{1985MIL}; 
(604)\,\cite{1987MIL}; 
(605)\,\cite{1988MIL}; 
(606)\,\cite{1989MIL}; 
(607)\,\cite{1997MIL}; 
(608)\,\cite{2001MIL}; 
(609)\,\cite{2003MIL}; 
(610)\,\cite{2004MIL}; 
(611)\,\cite{2008MIL}; 
(612)\,\cite{2010MIL}; 
(613)\,\cite{1984MIT}; 
(614)\,\cite{2009MIY}; 
(615)\,\cite{1993MON}; 
(616)\,\cite{1963MOR}; 
(617)\,\cite{2002MOR}; 
(618)\,\cite{2010MOR}; 
(619)\,\cite{2016MOR}; 
(620)\,\cite{1995MOS}; 
(621)\,\cite{2000MOS}; 
(622)\,\cite{2002MOS}; 
(623)\,\cite{2016MOS}; 
(624)\,\cite{2021MOT}; 
(625)\,\cite{2009MOU}; 
(626)\,\cite{2011MOU}; 
(627)\,\cite{2013MOU}; 
(628)\,\cite{2010MU}; 
(629)\,\cite{1979MUL}; 
(630)\,\cite{2005MUL}; 
(631)\,\cite{2018MUL}; 
(632)\,\cite{2019MUL}; 
(633)\,\cite{2020MUL}; 
(634)\,\cite{1986MUR}; 
(635)\,\cite{2003MURa}; 
(636)\,\cite{2003MURb}; 
(637)\,\cite{2005NAI}; 
(638)\,\cite{2009NAK}; 
(639)\,\cite{1989NAV}; 
(640)\,\cite{1981NEL}; 
(641)\,\cite{1982NEL}; 
(642)\,\cite{1989NEM}; 
(643)\,\cite{1988NES}; 
(644)\,\cite{1990NES}; 
(645)\,\cite{1998NGU}; 
(646)\,\cite{2000NGU}; 
(647)\,\cite{2001NGU}; 
(648)\,\cite{2003NGU}; 
(649)\,\cite{2006NGU}; 
(650)\,\cite{2011NGU}; 
(651)\,\cite{2012NGU}; 
(652)\,\cite{2014NGU}; 
(653)\,\cite{2018NGU}; 
(654)\,\cite{2019NGU}; 
(655)\,\cite{2020NGU}; 
(656)\,\cite{2000NIN}; 
(657)\,\cite{2003NIZ}; 
(658)\,\cite{2004NIZ}; 
(659)\,\cite{1981NOH}; 
(660)\,\cite{2018NUN}; 
(661)\,\cite{2020NUN}; 
(662)\,\cite{2015NUR}; 
(663)\,\cite{2017OCA}; 
(664)\,\cite{2004OEH}; 
(665)\,\cite{1990OHM}; 
(666)\,\cite{2016OLM}; 
(667)\,\cite{2013ONE}; 
(668)\,\cite{2014ONE}; 
(669)\,\cite{1996OPA}; 
(670)\,\cite{2003OSB}; 
(671)\,\cite{1994OYA}; 
(672)\,\cite{1979PAG}; 
(673)\,\cite{1988PAG}; 
(674)\,\cite{2009PAN}; 
(675)\,\cite{2020PAN}; 
(676)\,\cite{2023PAN}; 
(677)\,\cite{1993PAR}; 
(678)\,\cite{1994PAR}; 
(679)\,\cite{1999PARb}; 
(680)\,\cite{2001PAR}; 
(681)\,\cite{2004PAR}; 
(682)\,\cite{2006PAR}; 
(683)\,\cite{2008PAR}; 
(684)\,\cite{2011PAR}; 
(685)\,\cite{2013PAR}; 
(686)\,\cite{1995PAU}; 
(687)\,\cite{1996PAY}; 
(688)\,\cite{2019PEA}; 
(689)\,\cite{1996PEE}; 
(690)\,\cite{2003PEI}; 
(691)\,\cite{2018PEL}; 
(692)\,\cite{1999PEN}; 
(693)\,\cite{1969PER}; 
(694)\,\cite{1985PER}; 
(695)\,\cite{1988PER}; 
(696)\,\cite{1981PET}; 
(697)\,\cite{2002PET}; 
(698)\,\cite{2003PET}; 
(699)\,\cite{2009PEU}; 
(700)\,\cite{2013PEUa}; 
(701)\,\cite{2013PEUb}; 
(702)\,\cite{2018PEU}; 
(703)\,\cite{2016PHA}; 
(704)\,\cite{2019PHA}; 
(705)\,\cite{2020PHAa}; 
(706)\,\cite{2020PHAb}; 
(707)\,\cite{2020PHAc}; 
(708)\,\cite{2021PHA}; 
(709)\,\cite{1979PHI}; 
(710)\,\cite{1982PIT}; 
(711)\,\cite{1984PLA}; 
(712)\,\cite{2013PLA}; 
(713)\,\cite{2013POL}; 
(714)\,\cite{2000POP}; 
(715)\,\cite{2019POW}; 
(716)\,\cite{1959PSH}; 
(717)\,\cite{2020PUZ}; 
(718)\,\cite{2010QUA}; 
(719)\,\cite{2013RAG}; 
(720)\,\cite{2014RAG}; 
(721)\,\cite{2020RAJ}; 
(722)\,\cite{2013RAM}; 
(723)\,\cite{2007RAS}; 
(724)\,\cite{1977REI}; 
(725)\,\cite{2012REN}; 
(726)\,\cite{2012REY}; 
(727)\,\cite{1998RIMa}; 
(728)\,\cite{1998RIMb}; 
(729)\,\cite{1999RIM}; 
(730)\,\cite{2009RIS}; 
(731)\,\cite{2013RIS}.
}
\end{table}

\noindent
\begin{table}
\centering
\begin{tabular}{c}
\multicolumn{1}{c}{~~~~~~~~~~~~~~~~~~~~~~~~~~~~~~~~~~~~~~~~~~~~~~~~~~~~~~~~~~~~~~~~~~~~~~~~~~~~~~~~~~~~~~~~~~~~~~~~~~~~~~~~~~~~~~~~~~~~~~~~~~~~~~~~~~~~~~~~~~~~~~~~~~~~~~~~~~~~~~~~~~~~~~~~~~~~~~~~~~~~~~~~~~~~~~~~~~~~~~~~~~~~~} \\
\hline
\end{tabular}
\tablebib{
\tiny
(732)\,\cite{2014RIS}; 
(733)\,\cite{2015RIS}; 
(734)\,\cite{2014RIV}; 
(735)\,\cite{2017RIV}; 
(736)\,\cite{2006ROB}; 
(737)\,\cite{2023ROC}; 
(738)\,\cite{1994ROHa}; 
(739)\,\cite{1994ROHb}; 
(740)\,\cite{1995ROH}; 
(741)\,\cite{1973ROS}; 
(742)\,\cite{2005ROS}; 
(743)\,\cite{2010ROS}; 
(744)\,\cite{2011ROS}; 
(745)\,\cite{2018ROS}; 
(746)\,\cite{2020ROS}; 
(747)\,\cite{2001ROY}; 
(748)\,\cite{2011RUT}; 
(749)\,\cite{2017RYU}; 
(750)\,\cite{1990SAB}; 
(751)\,\cite{2012SAH}; 
(752)\,\cite{1986SAI}; 
(753)\,\cite{1988SAI}; 
(754)\,\cite{2013SAI}; 
(755)\,\cite{2018SAM}; 
(756)\,\cite{1987SAN}; 
(757)\,\cite{2017SAN}; 
(758)\,\cite{2019SAV}; 
(759)\,\cite{1958SCH}; 
(760)\,\cite{1969SCH}; 
(761)\,\cite{1973SCH}; 
(762)\,\cite{1985SCH}; 
(763)\,\cite{1994SCH}; 
(764)\,\cite{2019SEB}; 
(765)\,\cite{1994SEE}; 
(766)\,\cite{1979SEL}; 
(767)\,\cite{1996SEL}; 
(768)\,\cite{2008SEL}; 
(769)\,\cite{2017SEM}; 
(770)\,\cite{2018SEM}; 
(771)\,\cite{2002SEN}; 
(772)\,\cite{2005SENa}; 
(773)\,\cite{2005SENb}; 
(774)\,\cite{2006SENa}; 
(775)\,\cite{2006SENb}; 
(776)\,\cite{2007SENa}; 
(777)\,\cite{2007SENb}; 
(778)\,\cite{1995SER}; 
(779)\,\cite{2014SHA}; 
(780)\,\cite{2018SHA}; 
(781)\,\cite{2019SHAa}; 
(782)\,\cite{2019SHAb}; 
(783)\,\cite{2005SHE}; 
(784)\,\cite{1998SHI}; 
(785)\,\cite{2018SHI}; 
(786)\,\cite{2019SHI}; 
(787)\,\cite{2020SHI}; 
(788)\,\cite{1985SHU}; 
(789)\,\cite{2001SHUa}; 
(790)\,\cite{2001SHUb}; 
(791)\,\cite{2017SID}; 
(792)\,\cite{2025SIL}; 
(793)\,\cite{1972SIM}; 
(794)\,\cite{1988SIM}; 
(795)\,\cite{1992SIM}; 
(796)\,\cite{1993SIMb}; 
(797)\,\cite{1994SIM}; 
(798)\,\cite{1988SIN}; 
(799)\,\cite{2003SIV}; 
(800)\,\cite{2009SIV}; 
(801)\,\cite{2010SIVa}; 
(802)\,\cite{2010SIVb}; 
(803)\,\cite{2011SIVa}; 
(804)\,\cite{2011SIVb}; 
(805)\,\cite{2012SIV}; 
(806)\,\cite{2019SIV}; 
(807)\,\cite{2018SKO}; 
(808)\,\cite{1976SLA}; 
(809)\,\cite{1978SLA}; 
(810)\,\cite{1981SLA}; 
(811)\,\cite{1988SLA}; 
(812)\,\cite{1977SLE}; 
(813)\,\cite{2016SLE}; 
(814)\,\cite{2018SLEa}; 
(815)\,\cite{2018SLEb}; 
(816)\,\cite{GRI-Mech}; 
(817)\,\cite{2004SMI}; 
(818)\,\cite{2003SON}; 
(819)\,\cite{2017SON}; 
(820)\,\cite{2019SON}; 
(821)\,\cite{2010SOO}; 
(822)\,\cite{2021SOU}; 
(823)\,\cite{2004STR}; 
(824)\,\cite{2005SRI}; 
(825)\,\cite{1987STA}; 
(826)\,\cite{1988STE}; 
(827)\,\cite{1995STI}; 
(828)\,\cite{1998STI}; 
(829)\,\cite{1991STO}; 
(830)\,\cite{2000STO}; 
(831)\,\cite{2002STO}; 
(832)\,\cite{1994SU}; 
(833)\,\cite{1998SUM}; 
(834)\,\cite{2001SUN}; 
(835)\,\cite{2006SUN}; 
(836)\,\cite{2010SUN}; 
(837)\,\cite{2020SUN}; 
(838)\,\cite{2007SUZ}; 
(839)\,\cite{1978SWE}; 
(840)\,\cite{2003SWI}; 
(841)\,\cite{2011SWI}; 
(842)\,\cite{1984SZE}; 
(843)\,\cite{2010TAA}; 
(844)\,\cite{1979TAB}; 
(845)\,\cite{1977TAK}; 
(846)\,\cite{1999TAK}; 
(847)\,\cite{2006TAK}; 
(848)\,\cite{2007TAK}; 
(849)\,\cite{1996TAL}; 
(850)\,\cite{2002TAL}; 
(851)\,\cite{1979TAN}; 
(852)\,\cite{1980TAN}; 
(853)\,\cite{2001TAN}; 
(854)\,\cite{2007TANa}; 
(855)\,\cite{2007TANb}; 
(856)\,\cite{2008TAN}; 
(857)\,\cite{2008TAY}; 
(858)\,\cite{2020TER}; 
(859)\,\cite{1997THA}; 
(860)\,\cite{2001THI}; 
(861)\,\cite{2012THO}; 
(862)\,\cite{2004THW}; 
(863)\,\cite{2009TIA}; 
(864)\,\cite{1999TOK}; 
(865)\,\cite{2004TOK}; 
(866)\,\cite{2003TOM}; 
(867)\,\cite{2010TRA}; 
(868)\,\cite{2012TRA}; 
(869)\,\cite{2021TRA}; 
(870)\,\cite{1960TRE}; 
(871)\,\cite{1963TRE}; 
(872)\,\cite{2011TRE}; 
(873)\,\cite{2013TRE}; 
(874)\,\cite{2005TRO}; 
(875)\,\cite{2011TRO}; 
(876)\,\cite{2012TRO}; 
(877)\,\cite{1986TSA}; 
(878)\,\cite{1987TSA}; 
(879)\,\cite{1988TSA}; 
(880)\,\cite{1989TSA}; 
(881)\,\cite{1990TSAa}; 
(882)\,\cite{1990TSAb}; 
(883)\,\cite{1991TSAa}; 
(884)\,\cite{1991TSAb}; 
(885)\,\cite{1992TSAa}; 
(886)\,\cite{1992TSAb}; 
(887)\,\cite{2004TSA}; 
(888)\,\cite{1994TSU}; 
(889)\,\cite{1996TSU}; 
(890)\,\cite{1997TSU}; 
(891)\,\cite{1975TUL}; 
(892)\,\cite{1995TWA}; 
(893)\,\cite{1998TYN}; 
(894)\,\cite{1997UPA}; 
(895)\,\cite{1995VAG}; 
(896)\,\cite{1996VAG}; 
(897)\,\cite{2001VAG}; 
(898)\,\cite{2020VAG}; 
(899)\,\cite{2001VAK}; 
(900)\,\cite{2003VAK}; 
(901)\,\cite{1981VAN}; 
(902)\,\cite{2010VAN}; 
(903)\,\cite{2016VAN}; 
(904)\,\cite{2016VAR}; 
(905)\,\cite{2005VAS}; 
(906)\,\cite{2010VASa}; 
(907)\,\cite{2010VASb}; 
(908)\,\cite{2002VER}; 
(909)\,\cite{2003VER}; 
(910)\,\cite{1982VEY}; 
(911)\,\cite{2017VIC}; 
(912)\,\cite{2017VID}; 
(913)\,\cite{2008VRA}; 
(914)\,\cite{2012VUI}; 
(915)\,\cite{2015WAK}; 
(916)\,\cite{1989WAL}; 
(917)\,\cite{1998WAL}; 
(918)\,\cite{1987WAN}; 
(919)\,\cite{1991WAN}; 
(920)\,\cite{1997WAN}; 
(921)\,\cite{2001WAN}; 
(922)\,\cite{2003WAN}; 
(923)\,\cite{2005WANa}; 
(924)\,\cite{2005WANb}; 
(925)\,\cite{2006WAN}; 
(926)\,\cite{2015WAN}; 
(927)\,\cite{2017WAN}; 
(928)\,\cite{2018WAN}; 
(929)\,\cite{2019WANa}; 
(930)\,\cite{2020WAN}; 
(931)\,\cite{1984WAR}; 
(932)\,\cite{1993WAT}; 
(933)\,\cite{1991WAY}; 
(934)\,\cite{1988WEI}; 
(935)\,\cite{2004WEI}; 
(936)\,\cite{1994WEN}; 
(937)\,\cite{1975WES}; 
(938)\,\cite{1980WES}; 
(939)\,\cite{1989WES}; 
(940)\,\cite{2019WES}; 
(941)\,\cite{1967WIL}; 
(942)\,\cite{1972WIL}; 
(943)\,\cite{1998WIL}; 
(944)\,\cite{2012WIL}; 
(945)\,\cite{1996WOI}; 
(946)\,\cite{1971WOO}; 
(947)\,\cite{1994WOO}; 
(948)\,\cite{1996WOO}; 
(949)\,\cite{2007WU}; 
(950)\,\cite{2018WU}; 
(951)\,\cite{2019WU}; 
(952)\,\cite{2005XIE}; 
(953)\,\cite{2006XIE}; 
(954)\,\cite{2023XIE}; 
(955)\,\cite{1996XIN}; 
(956)\,\cite{2014XIO}; 
(957)\,\cite{1997XU}; 
(958)\,\cite{1998XU}; 
(959)\,\cite{1999XUa}; 
(960)\,\cite{1999XUb}; 
(961)\,\cite{2004XUa}; 
(962)\,\cite{2004XUb}; 
(963)\,\cite{2007XUa}; 
(964)\,\cite{2007XUb}; 
(965)\,\cite{2009XU}; 
(966)\,\cite{2010XUa}; 
(967)\,\cite{2010XUb}; 
(968)\,\cite{2011XU}; 
(969)\,\cite{2020XU}; 
(970)\,\cite{2022XU}; 
(971)\,\cite{1992YANa}; 
(972)\,\cite{1992YANb}; 
(973)\,\cite{1994YAN}; 
(974)\,\cite{2004YAN}; 
(975)\,\cite{2018YAN}; 
(976)\,\cite{2020YAN}; 
(977)\,\cite{2008YAS}; 
(978)\,\cite{2009YAS}; 
(979)\,\cite{2010YEL}; 
(980)\,\cite{1989YET}; 
(981)\,\cite{2006YIL}; 
(982)\,\cite{2020YON}; 
(983)\,\cite{1958YOU}; 
(984)\,\cite{1988ZAB}; 
(985)\,\cite{1989ZABa}; 
(986)\,\cite{1989ZABb}; 
(987)\,\cite{1989ZABc}; 
(988)\,\cite{2009ZAD}; 
(989)\,\cite{2011ZAD}; 
(990)\,\cite{2015ZAD}; 
(991)\,\cite{2017ZAD}; 
(992)\,\cite{2016ZAH}; 
(993)\,\cite{2009ZAN}; 
(994)\,\cite{2018ZAN}; 
(995)\,\cite{2020ZARa}; 
(996)\,\cite{2020ZARb}; 
(997)\,\cite{1988ZEL}; 
(998)\,\cite{2016ZEN}; 
(999)\,\cite{2002ZHA}; 
(1000)\,\cite{2004ZHA}; 
(1001)\,\cite{2005ZHAa}; 
(1002)\,\cite{2005ZHAb}; 
(1003)\,\cite{2005ZHAc}; 
(1004)\,\cite{2006ZHA}; 
(1005)\,\cite{2007ZHA}; 
(1006)\,\cite{2008ZHA}; 
(1007)\,\cite{2011ZHA}; 
(1008)\,\cite{2012ZHA}; 
(1009)\,\cite{2014ZHA}; 
(1010)\,\cite{2016ZHA}; 
(1011)\,\cite{2018ZHAa}; 
(1012)\,\cite{2018ZHAc}; 
(1013)\,\cite{2018ZHAd}; 
(1014)\,\cite{2019ZHAa}; 
(1015)\,\cite{2019ZHAc}; 
(1016)\,\cite{2019ZHAb}; 
(1017)\,\cite{2019ZHAd}; 
(1018)\,\cite{1998ZHO}; 
(1019)\,\cite{2009ZHO}; 
(1020)\,\cite{2013ZHO}; 
(1021)\,\cite{2017ZHO}; 
(1022)\,\cite{2019ZHO}; 
(1023)\,\cite{2008ZHU}; 
(1024)\,\cite{2012ZHU}; 
(1025)\,\cite{2020ZHU}.
}
\end{table}

\end{appendix}

\end{document}